\documentclass[namedreferences,hyperref,optionalrh,solaromanenum]{spr-sola}

\usepackage{amsmath}
\usepackage{lscape}
\usepackage{tabularx}
\usepackage{rotating}
\usepackage{graphicx}        % For eps figures, newer & more powerfull
\usepackage{amssymb}        % useful mathematical symbols
\usepackage{color}           % For color text: \color command
\renewcommand{\vec}[1]{{\mathbfit #1}}

\newcommand{\pder}[2]{\frac{\partial #1}{\partial #2} }
\newcommand{\grad}{ {\bf \nabla } }
\newcommand{\divv}{ {\bf \nabla} \cdot }
\newcommand{\curl}{ {\bf \nabla} \times}

\begin{document}
\begin{article}
\begin{opening}

%\title{Simulations of standing and reflected slow-mode waves in flaring coronal loops with a 2D arcade model }
\title{Exploring Standing and Reflected Slow-mode Waves in Flaring Coronal Loops: A Parametric Study Using 2.5D MHD Modeling}

\author[addressref={aff1,aff2},corref,email={tongjiang.wang@nasa.gov}]{\inits{T.J.}\fnm{Tongjiang}~\lnm{Wang}\orcid{3-0053-1146}}
\author[addressref={aff1,aff2}]{\inits{L.}\fnm{Leon}~\lnm{Ofman}\orcid{3-0602-6693}}
\author[addressref=aff3]{\inits{S.J.}\fnm{Stephen J.}~\lnm{Bradshaw}\orcid{2-3300-6041}}

%email={ofman@cua.edu}
%email={stephen.bradshaw@rice.edu}
%\author[addressref={aff1,aff2}],{\inits{L.}\fnm{Leon}~\lnm{Ofman}}
%\author[addressref=aff3],{\inits{S.J.}\fnm{Stephen}~\lnm{Bradshaw}}
\address[id=aff1]{Department of Physics, Catholic University of America, 620 Michigan Avenue NE, Washington, DC 20064, USA}
\address[id=aff2]{NASA Goddard Space Flight Center, Code 671, Greenbelt, MD 20770, USA}
\address[id=aff3]{Department of Physics and Astronomy, Rice University, Houston, TX 77005, USA}

%\author{\inits{}\fnm{}~\lnm{}\orcid{}}
%   NOTE:  Just one corresponding author [corref]

%%%%%%%%%%%%%%%%%%%%%%%%%%%%%%%%%%%%%%%%%%%%%%%%%%%^M
%% Runningheads^M
%^M
 \runningauthor{Wang et al.}
 \runningtitle{A parametric study of slow-mode waves in flaring coronal loops}
%
%%% Abstract (average 150 words, maximum 300 words)
\begin{abstract}
	Recent observations of reflected propagating and standing slow-mode waves in hot flaring coronal loops have spurred our investigation into their underlying excitation and damping mechanisms. To understand these processes, we conduct 2.5D magnetohydrodynamic (MHD) simulations using an arcade active region model that includes a hot and dense loop. Our simulations allow for in-depth parametric investigations complementing and expanding our previous 3D MHD modeling results. We excite these waves using a large-amplitude, flow pulse applied at one footpoint of the loop in two distinct models as motivated by observations from the {\it Solar Dynamics Observatory/Atmospheric Imaging Assembly} (SDO/AIA). The first model (Model 1) incorporates classical compressive viscosity coefficient, while the second model (Model 2) adopts a 10-times enhanced viscosity coefficient. We obtain the following major results: (1) Our 2.5D MHD simulations reinforce previous conclusions derived from 1D and 3D MHD models that significantly enhanced viscosity is crucial for the rapid excitation of standing slow waves with damping times consistent with observations by \citet{wan15}. (2) We uncover that nonlinearity in Model 1 delays the conversion of a reflected propagating wave into a standing wave. In contrast, Model 2 exhibits a much weak influence of nonlinearity on the excitation time of standing waves, thanks to the suppression of these effects by enhanced viscosity. (3) Our results reveal that the transverse temperature structure holds more influence on wave behavior than the density structure. In Model 1, increased loop temperature contrast significantly enhances wave trapping within the structure, mitigating the impact of temperature-dependent viscous damping. Conversely, in Model 2, the impact of temperature structure on wave behavior weakens in comparison to the effect of viscosity. (4) Model 1 displays evident nonlinear coupling to the fast and kink magnetoacoustic waves and pronounced wave leakage into the corona. Model 2 exhibits significantly weaker effects in this regard. Analyzing three observed wave events by SDO/AIA aligns with Model 2 predictions, providing further support for the substantial viscosity increase. Our 2.5D study unravels the complex interplay of wave-flow phenomena and nonlinear processes in coronal loops, extending our previous 1D modeling results to incorporate more realistic loop geometry. This provides insights into scenarios where 3D effects may be neglected, thereby enhancing our understanding of the intricate dynamics of the solar corona. 
\end{abstract}

%%%%%%%%%%%%%%%%%%%%%%%%%%%%%%%%%%%%%%%%%%%%%%%%%%%^M
%% Keywords^M
%^M

\end{opening}

%-------------------------------------------------^M
%%%%%%%%%%%%%%%%%%%%%%%%%%%%%%%%%%%%%%%%%%%%%%%%%%%^M
%% Sections^M
%^M
\section{Introduction}
The phenomenon of slow magnetoacoustic waves in hot loops of active region (AR) coronae has been an active research topic in observation and theory since its discovery \citep[see reviews by][]{wan11,wan21}. This is because, through a technique known as coronal seismology \citep[e.g., reviews by][]{nak05,wan16,nak20}, important information about the coronal (loop) structure can be extracted from the observed wave properties, which would otherwise be impossible or challenging to obtain. This information includes the magnetic-field strength \citep{wan07,nis17}, transport coefficients \citep{wan15,wan19,kolot2022}, heating functions \citep{real19,kol20,kol22}, and the potential role of slow wave dissipation in coronal heating \citep{xia22}. 

Impulsively-generated Doppler velocity oscillations were first discovered by the {\it Solar and Heliospheric Observatory/Solar Ultraviolet Measurements of Emitted Radiation} (SOHO/SUMER) spectrometer in flare emission lines, mainly Fe\,{\sc xix} and Fe\,{\sc xxi}, with formation temperatures above 6 MK \citep[e.g.,][]{wan02,wan03a,wan07}. These oscillations were mostly interpreted as the fundamental standing slow-mode waves based on the measured phase speed, close to the sound speed at the loop’s temperature, and the presence of a quarter-period phase shift between velocity and intensity disturbances \citep[e.g.,][]{ofm02,wan03b}. Similar Doppler velocity oscillations detected in the S\,{\sc xv} and Ca\,{\sc xix} lines by {\it Yohkoh/Bragg Crystal Spectrometer} (Yohkoh/BCS) were also interpreted as the standing slow-mode waves \citep{mari05,mari06}. 

The launch of SDO has led to the discovery of longitudinal intensity oscillations by AIA in high-temperature EUV emission channels (94 \AA\ of 7 MK and 131 \AA\ of 11 MK), characterized by apparent sloshing motions of the disturbance between the two footpoints of a hot flaring loop \citep{kum13,wan15}. Similar intensity oscillations were also observed in hot loops by {\it Hinode/X-Ray Telescope} (Hinode/XRT) \citep{mand16}. The longitudinal oscillations from imaging observations, showing wave and plasma properties consistent with those of the SUMER oscillations, have been mostly interpreted as the slow-mode waves in a reflected propagating mode \citep{kum13,kum15,fang15,mand16,nis17,xia22}. However, evidence for a transition from an initial reflected propagating wave into the standing wave pattern has been revealed during the loop-cooling phase \citep{kri21}. In contrast, \citet{wan15} detected one case showing that the initial disturbance quickly forms a standing wave after only one reflection, in agreement with what was detected by SUMER \citep[e.g.,][]{wan03b}. 

Both standing and reflected propagating slow-mode waves in hot coronal loops are observed to decay quickly, typically lasting for a couple of periods \citep[e.g.,][]{wan21}. Statistical studies have shown that the measured decay times and wave periods are comparable and follow a near-linear scaling relationship \citep{wan03a,wan05,mari06,nak19}. Thermal conduction was suggested to be the dominant damping mechanism for slow waves in typical hot coronal loops, based on linear theories \citep{dem03,pan06,pras21} and nonlinear MHD simulations \citep{ofm02,mend04,brad08,fang15}. Many other damping mechanisms have been proposed, including nonlinearity effects \citep{verw08,rud13}, wave leakage into the corona due to curvature and mode coupling \citep{selw07,ogr09, ofm12}, and wave-caused heating-cooling misbalance \citep{kums16,nak17,kolot19}. The near-linear scaling between damping times and oscillation periods was interpreted as the combined effect of thermal conduction and compressive viscosity \citep{ofm02,wan21}. 

\citet{pras21} demonstrated that by considering the joint effect of thermal conduction, compressive viscosity, and heating-cooling misbalance, the theoretically predicted values of damping time and wave period can be fitted to a function composed of two power-law scalings (corresponding to short- and long-period domains, respectively), which better matches the SUMER observations. From the linear theory, including effects of wave-induced thermal misbalance, \citet{kol22} derived a new relationship between damping time and wave period, which suggests a more generic seismological technique to constrain the coronal heating function.  

The rapid excitation of standing slow-mode waves in hot coronal loops is a challenging problem that has attracted many theoretical studies \citep{wan21}. Observations indicate that both SUMER and AIA loop oscillations are mostly triggered by small- or micro-flares at one footpoint of a hot loop system. Various MHD models have shown that impulsive heating typically excites the reflected propagating slow-mode waves (also called sloshing oscillations), which undergo several reflections between two footpoints of the loop to form a standing wave \citep{selw05,tar05,ogr07,ogr09,fang15,real16,wan18,ofm22}. Strongly asymmetric heating (i.e., at one footpoint) favors the formation of the fundamental mode, while symmetric heating results in the second harmonic \citep[e.g.,][]{selw05}. 

To address the issue of rapid excitation (within about one oscillation period) of the fundamental standing slow-mode wave by single footpoint heating, various mechanisms have been proposed: (1) launching a long pulse with a duration comparable to the wave period using the 1D HD model \citep{tar05,tar07}; (2) launching a broad velocity pulse at one footpoint of a curved loop that leads to the interaction of a primary perturbation propagating inside the loop with a second slow pulse excited at the remote footpoint by the external fast wave using 2D and 3D  modeling \citep{selw07,selw09};  (3) injecting a steady flow or flow pulse into the loop to excite the slow waves that can quickly form the standing mode due to significant wave leakage into the corona related to coupling with fast-mode waves using a 3D resistive MHD model \citep{ofm12,ofm22}; (4) significantly enhanced compressive viscosity that leads to efficient dissipation of high-frequency components in a flow pulse based on a nonlinear 1D MHD model \citep{wan18,wan19}.

Specifically, the work by \citet{selw09} delves into the excitation and damping of slow standing waves within a dense loop, using a 3D resistive MHD model in an isothermal atmosphere while neglecting viscosity and gravity. They arrived at a similar conclusion to that of \citet{selw07}, yet they observed a distinct phenomenon: the excitation of slow standing waves through a footpoint pulse becomes more efficient in the 3D case compared to the 2D case, primarily due to the heightened effectiveness of wave leakage out of the loop in a 3D geometry. One notable factor contributing to this pronounced wave leakage is the presence of plasma-$\beta$ exceeding 1 at the loop's apex within their model conditions, a condition that likely happens in hot and dense postflare loops. Nevertheless, measurements by \citet{wan07} showed that the typical values of plasma-$\beta$ in hot oscillating loops remain on the order of 0.1.   

\citet{ofm12} studied impulsively generated waves and flows in hot coronal loops using 3D MHD modeling in the isothermal case without considering a pre-existing density structure, and found that the impulsive onset of steady flows can excite damped magnetoacoustic waves that initially propagate along the loop and form a standing wave quickly. They suggested that the quick damping and fast formation of standing waves could be attributed to large wave leakage due to nonlinear coupling of slow waves with fast-mode waves. \citet{ofm22} studied the excitation and damping of slow-mode waves in a hot and dense coronal loop using a 3D MHD model including the effects of compressive viscosity along the magnetic field. They found that there is a small effect of viscosity dissipation in a 6 MK hot loop in the warm (2 MK) coronal background compared to significant wave leakage due to mode coupling, whereas the more effects of viscous dissipation in hotter ($\sim$10 MK) coronal loops in the hot (7 MK) coronal background. It needs to be pointed out that the initial hot and dense loop is not exactly in a stable equilibrium in their modeling. The gradual evolution of the loop in transverse structure may increase the leakage of waves into the corona.

By analyzing an AIA loop oscillation event, \citet{wan15} found evidence for the suppression of thermal conduction, supported by the near-inphase relationship between temperature and density variations and the measured polytropic index having a value close to 5/3. They also suggested that classical compressive viscosity needs to be enhanced by an order of magnitude to account for the rapid decay of observed waves. In a study by \citet{pras22}, the role of non-ideal dissipation in the phase shift and polytropic index of standing slow waves was investigated based on linear theory. The results showed that, with consideration of thermal misbalance, the conclusion of suppressed thermal conduction and enhanced viscosity obtained in \citet{wan15} remains valid. 

Further research by \citet{wan19} involved numerical parametric studies to refine the values of transport coefficients determined in \citet{wan15}. Additionally, \citet{wan18} utilized 1D nonlinear MHD simulations to study the role of modified transport coefficients in the excitation of standing slow waves. They demonstrated that the model with seismology-determined transport coefficients can self-consistently produce the standing slow mode wave as quickly (within one period) as observed, In contrast, the model with classical transport coefficients produces an initial reflected propagating wave that requires many footpoint reflections to form a standing wave.

Recently, \citet{ofm22} investigated the excitation and damping of slow waves in hot coronal loops using nonlinear 3D MHD models featuring a bipolar magnetic geometry. They showed the significant role of wave leakage in inducing damping, despite the enhanced trapping effect brought about by the transverse temperature structure within the loop. Nevertheless, their analyses remained qualitative, constrained by the inherent limitations of 3D MHD modeling in terms of computational accuracy and expense. 

In this study, we employ a 2.5D arcade loop model to explore the influence of modified transport coefficients on the excitation of the slow-mode wave and the impact of transverse structuring in temperature and density within the loop on the phenomena of wave trapping/leakage. Utilizing 2.5D MHD models affords us the opportunity to conduct numerous long-duration numerical simulations characterized by high resolutions and low computational costs, enabling a thorough parametric investigation without sacrificing generality compared to 3D counterparts. Section~\ref{sct:mdl} outlines the numerical model, initial setup, and boundary conditions. Section~\ref{sct:rst} presents the modeling results and quantitative analyses of the effects of enhanced compressive viscosity, as well as other factors including thermal conduction, nonlinearity, and transverse structuring on wave damping. Comparisons of the simulations with some observed events regarding mode coupling and wave leakage are also included in this section. Finally, we provide the discussion and conclusions in Section~\ref{sct:dc}.

\section{2.5D Hot Arcade Loop Modeling}
\label{sct:mdl}
\subsection{MHD Model}
To model plasma dynamics in a 2.5D arcade active region we solve the nonlinear, resistive 3D MHD equations using a 3D code NLRAT (run in 2D mode). The simulation includes effects of gravity, compressive viscosity, heat conduction, and optically thin radiation \citep[see][in detail]{prov18,ofm18,ofm22}. The MHD equations are presented in a flux-conservative and dimensionless form:

\begin{equation}
 \pder{\rho}{t} + \divv ( \rho \vec{V} ) =0,  \label{eq:cn} 
\end{equation}
\begin{equation}
 \pder{(\rho \vec{V})}{t} + \divv \left[ \rho \vec{V}\vec{V} + \left( E_u \, p + \frac{\vec{B} \cdot \vec{B}}{2} \right) \vec{I} - \vec{B} \vec{B} \right] = \frac{1}{F_r} \rho \vec{F_g} +\vec{F_v}, \label{eq:mm}
\end{equation}
\begin{multline}
  \pder{(\rho E)}{t} + \divv \left[\vec{V} \left(\rho E + E_u \, p + \frac{\vec{B} \cdot \vec{B}}{2} \right) - \vec{B}(\vec{B} \cdot \vec{V}) + \frac{1}{S} \curl \vec{B}\times\vec{B} + \vec{V}\cdot\mathbf{\Pi} \right] \\
 =\frac{1}{F_r} \rho \vec{F_g} \cdot \vec{V} + \grad_\| (\kappa_\|\grad_{\|}T) - Q_{\rm rad} + H_{\rm in},
	 \label{eq:en}
\end{multline}
\begin{equation}
  \pder{\vec{B}}{t} = \curl (\vec{V}\times\vec{B}) + \frac{1}{S} \nabla^2 \vec{B}. \label{eq:ind}
\end{equation}
In the above equations, the total energy density is represented as $\rho{E}=\frac{E_u p}{(\gamma-1)}+\frac{\rho{V^2}}{2}+\frac{B^2}{2}$, where $p$, $\rho$, $V$, and $B$ correspond to the dimensionless pressure, density, velocity, and magnetic field, respectively. The gravity term is denoted as $\vec{F_g}=-\frac{1}{(10+z-z_{\rm min})^2}\vec{e_z}$, the gradient parallel to the magnetic field is defined as $\grad_\|=\frac{1}{|B|}\vec{B}\cdot\grad$ and is considered as a scalar operator. The $B$-parallel heat conductivity is given by $\kappa_\|=7.8\times10^{-7} T^{5/2}$ erg~cm$^{-1}$s$^{-1}$K$^{-1}$ based on \citet{spit53}. 

The viscous stress tensor is denoted as $\mathbf{\Pi}$, and the viscous force is represented as $\vec{F_v}=-\divv \mathbf{\Pi}$. Additionally, a term on the left-hand side of the energy equation, denoted as $S_v=-\divv (\vec{V}\cdot\mathbf{\Pi})$, represents the total energy flux transferred by viscous forces \citep[see Equation 6.33 in][]{brag65}. This term is crucial for understanding the energy dissipation and transfer processes due to viscosity in the plasma. $S_v$ can be decomposed into two parts: the volumetric heating rate ($Q_v=-\mathbf{\Pi}\cdot\grad\vec{V}$) due to viscous dissipation and the rate of work done ($W_v=\vec{V}\cdot\vec{F_v}$) by the viscous forces on the plasma. The presence of viscous effects can have significant implications for the dynamics and energetics of the coronal plasma, particularly in conditions where the magnetic field is strong (e.g., in active region loops) and the viscosity becomes highly nonisotropic. The dominant terms in the viscous stress tensor, corresponding to the compressive viscosity, can be expressed as follows \citep{brag65,holl86,craig07},  
\begin{equation}
  \Pi_{ij} = 3\eta_0 \left(\frac{\delta_{ij}}{3}-\frac{B_i B_j}{B^2} \right) \left(\frac{\vec{B}\cdot\vec{B}\cdot\grad\vec{V}}{B^2}-\frac{\divv\vec{V}}{3} \right),  \label{eq:vis}
\end{equation}
where $\eta_0=10^{-16} T^{5/2}$ g~cm$^{-1}$s$^{-1}$ is the Braginskii compressive viscosity coefficient. From Equation~\ref{eq:vis}, it follows that the viscous heating rate is positive definite and given by $Q_v=3\eta_0 (B^{-2}\vec{B}\cdot\vec{B}\cdot\grad\vec{V}-\divv\vec{V}/3)^2$ \citep[see][]{holl86}. In the 1D case with $\vec{B}=B\vec{z}$, this simplifies to $Q_v=(4/3)\eta_0(\partial{V_z}/\partial{z})^2$. In a strongly magnetized plasma, viscous heating refers to an irreversible process where the energy of the plasma is converted into internal energy (heat) due to the action of anisotropic viscous forces. On the other hand, the term $W_v$ can either add mechanical energy to the plasma (positive work) or dissipate its kinetic energy into other forms (negative work). The net effect depends on the relative orientation of the viscous force and velocity vector.

In the present study, we omit the radiative loss term $Q_{\rm rad}$ and the empirical heating function $H_{\rm in}$, which is often employed to balance the radiative cooling. Instead, we adopt an empirical polytropic index value of $\gamma$=1.05 \citep[e.g.][]{link99}. This choice reduces the impact of the source terms in the energy equation, as the plasma is nearly isothermal with an empirical value of $\gamma$ close to unity. 

The normalization of the MHD equations results in the dimensionless parameters: the Euler number $E_u ={4\pi p_0}/{B_0^2} ={C_{s0}^2}/{(\gamma V_{A0}^2)}={\beta_0}/{2}$, the Froude number $F_r =  {V_{A0}^2 L_0}/(G M_s)$, and the Lundquist number $S = {L_0 V_{A0}}/{\eta}$. Here, $L_0$ represents the length scale defined as $L_0=0.1 R_s$ (where $R_s$ is the solar radius), $B_0$ is the normalizing magnetic field magnitude, $\rho_0$ is the normalizing density, and $p_0$ is the normalizing pressure in the corona. Additionally, $V_{A0}=B_0/(4\pi\rho_0)^{1/2}$ is the normalizing Alfv\'{e}n speed, $C_{s0}=(\gamma{p_0}/\rho_0)^{1/2}$ is the characteristic sound speed, $G$ denotes the gravitational constant, $M_s$ represents the solar mass, and $\eta$ is the resistivity. 

The physical parameters used in the present model are summarized in Table~\ref{tab:para}. We have set $S = 10^4$ in this study, which implies that the resistivity in the model is much higher than solar resistivity due to numerical limitations. However, despite this choice, the resistivity has a negligible effect on the slow magnetoacoustic waves.

\begin{table}
\caption{ Physical parameters used in the model.}
\label{tab:para}
\begin{tabularx}{\textwidth}{lclc} 
  \hline
Quantities & Values & Quantities & Values \\
  \hline
Length scale ($L_0$) & 0.1$R_s$ & Loop length ($L$) & 187$\pm$54 Mm\\
Magnetic field ($B_0$) & 100 G & Loop width ($w$) & 10 Mm\\
Temperature ($T_0$) & 7 MK & Pulse amplitude ($A_0$) & 0.1 {\rm or} 0.01\\
Number density ($n_0$) & $10^9$ cm$^{-3}$ & Pulse duration ($\Delta{t}$) & 30 $\tau_A$\\
	Alfv\'{e}n speed ($V_{A0}$) & 6898 km~s$^{-1}$ & $\kappa_{\|}(T_0)$ & \makebox[3.1cm][r]{1.01$\times10^{11}$ erg\,{(cm\,s\,K)}$^{-1}$}\\
Alfv\'{e}n time ($\tau_A$) & 10.1 s & $\eta_0(T_0)$ & 12.9 g~{\rm (cm s)}$^{-1}$ \\
Sound speed ($C_{s0}$) & 348 km~s$^{-1}$ & Plasma-$\beta_0$ & 0.0049\\
Grav. scale height ($H_0$) & 427 Mm & Polytropic index ($\gamma$) & 1.05\\
Euler number ($E_u$) & $2.43\times10^{-3}$ & Froude number ($F_r$) & 25.1 \\
  \hline
\end{tabularx}
\end{table}
%\multicolumn{1}{l}{##}

\begin{table}
\caption{Parameters used in the two types of numerical models for the various cases. Model 1 employs the classical values ($\eta_c$) of the compressive viscosity coefficient $\eta_0$, while model 2 uses the ten times enhanced values (10$\eta_c$). Cases \#A-F are the control numerical experiments that differ from case \# only in one parameter, where \# represents model 1 or 2. The classical values of the $B$-parallel thermal conductivity $\kappa_\|$ are denoted as $\kappa_c$. The loop density and temperature contrasts at the footpoint are represented by $\chi_\rho$ and $\chi_T$, respectively. $A_0$ is the normalized injection velocity amplitude. The term $S_v$ is assigned the value `Y' (`N') to indicate the inclusion (exclusion) of the viscous term in the energy equation. }
\label{tab:cases}
\begin{tabularx}{\textwidth}{rcccccccrcccccc} 
  \hline
 Case & $\eta_0$ & $\kappa_\|$ & $\chi_\rho$ & $\chi_T$ & $A_0$ & $S_v$ && Case & $\eta_0$ & $\kappa_\|$ & $\chi_\rho$ & $\chi_T$ & $A_0$ & $S_v$ \\
  \hline
1 & $\eta_c$ & $\kappa_c$ & 1.5 & 1.5 & 0.1 & Y && 2 & 10$\eta_c$ & $\kappa_c$ & 1.5 & 1.5 & 0.1 & Y \\
1A & $\eta_c$ & 0 & 1.5 & 1.5 & 0.1 & Y && 2A & 10$\eta_c$ & 0 & 1.5 & 1.5 & 0.1 & Y \\
1B & $\eta_c$ & $\kappa_c$ & 1.5 & 1.5 & 0.01 & Y && 2B & 10$\eta_c$ & $\kappa_c$ & 1.5 & 1.5 & 0.01 & Y \\
1C & $\eta_c$ & $\kappa_c$ & 1.0 & 1.5 & 0.1 & Y && 2C & 10$\eta_c$ & $\kappa_c$ & 1.0 & 1.5 & 0.1 & Y \\
1D & $\eta_c$ & $\kappa_c$ & 1.5 & 1.0 & 0.1 & Y && 2D & 10$\eta_c$ & $\kappa_c$ & 1.5 & 1.0 & 0.1 & Y \\
1E & $\eta_c$ & $\kappa_c$ & 1.5 & 2.0 & 0.1 & Y && 2E & 10$\eta_c$ & $\kappa_c$ & 1.5 & 2.0 & 0.1 & Y \\
1F & $\eta_c$ & $\kappa_c$ & 1.5 & 1.5 & 0.1 & N && 2F & 10$\eta_c$ & $\kappa_c$ & 1.5 & 1.5 & 0.1 & N \\
% \cline{1-7} \cline{9-15}\\
% \multicolumn{7}{c}{Model 1} & & \multicolumn{7}{c}{Model 2} \\
 \hline
\end{tabularx}
\end{table}

\begin{figure}
\centerline{\includegraphics[width=1.0\textwidth,clip=]{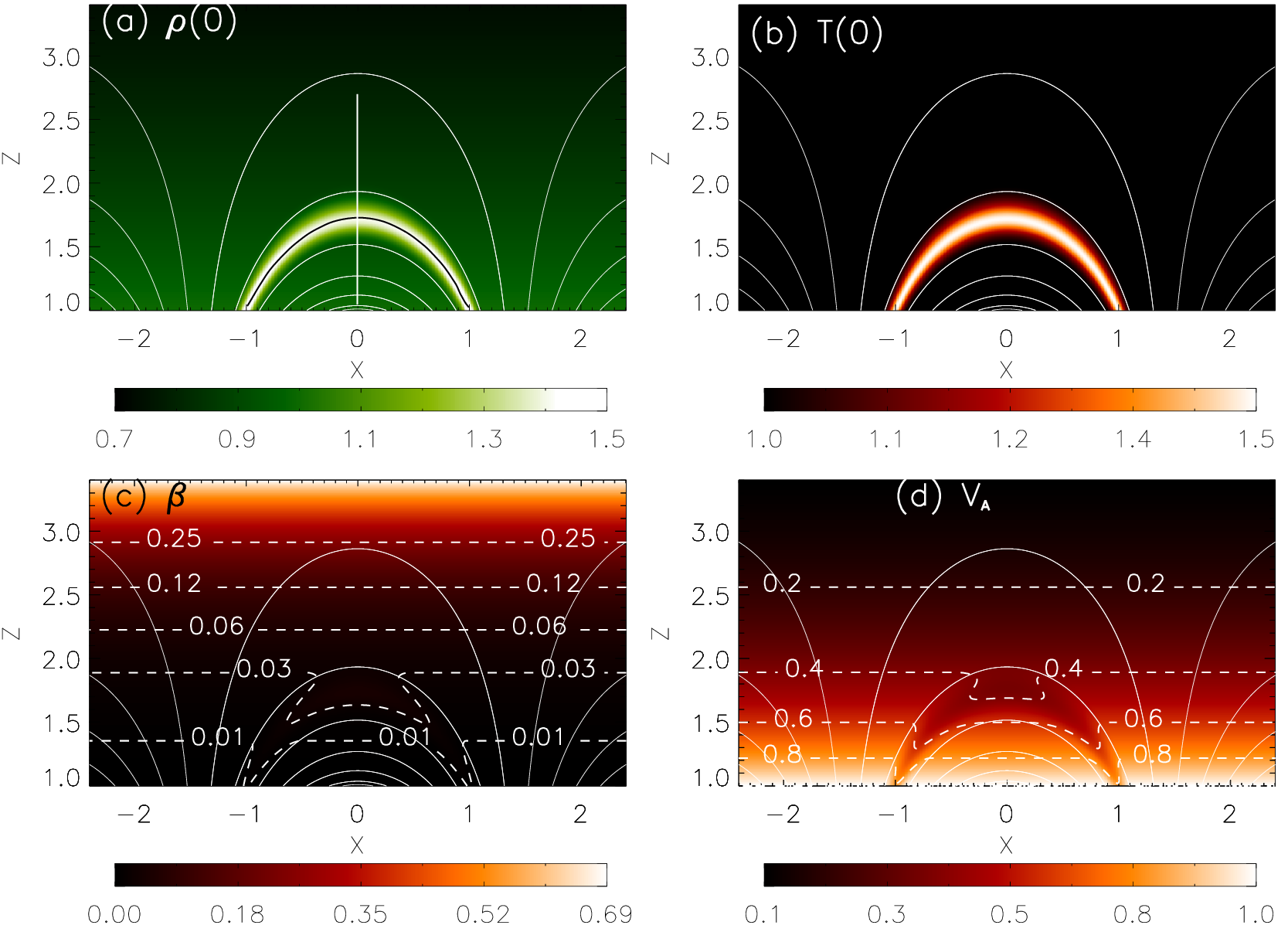}}
\caption{Initial configuration of a 2D arcade loop model in polytropic equilibrium. (a) Density distribution ($\rho(0)$; color scale in units of $n_0 m_p$, where $n_0=10^9$ cm$^{-3}$ and $m_p$ is the proton mass). (b) Temperature distribution ($T(0)$; color scale in units of $T_0$=7 MK). (c) Plasma $\beta$, overlaid with the contours (dashed lines) showing the values of $\beta$=0.01, 0.03, 0.06, 0.12, 0.25. (d) Local Alfv\'{e}n speed ($V_A$; color scale in units of $V_{A0}$=6898 km~s$^{-1}$), overlaid with the contours (dashed lines) showing the levels of $V_A$=0.2, 0.4, 0.6, 0.8. White solid curves in each panel depict magnetic field lines for a potential field model. In panel (a), a black curve indicates a cut along the loop model used to generate time distance plots presented in Figure~\ref{fig:tdis}, and a white vertical line indicates a cut across the loop's apex used to generate time distance plots presented in Figure~\ref{fig:vtds}.}
\label{fig:init}
\end{figure}

\subsection{Initial Setup  }
\label{sct:isp}
A coronal arcade model in Cartesian geometry is initialized with the potential magnetic field, as described in numerous previous studies \citep[e.g.,][]{oli93,selw05,selw07}. In this model, the magnetic field components in dimensionless form are represented as follows:
\begin{eqnarray}
 B_x&=&-{\rm cos}(x/\Lambda_B){\rm e}^{-(z-z_{\rm min})/\Lambda_B}, \\
 B_y&=&0, \\
 B_z&=&{\rm sin}(x/\Lambda_B){\rm e}^{-(z-z_{\rm min})/\Lambda_B},
\end{eqnarray}
where $\Lambda_B=2L_w/\pi$ represents the magnetic scale height, with $L_w$ denoting the half-width of the arcade (for which we take $L_w=1.43$ in this study). It is evident that the normalizing magnetic field $B_0$ corresponds to the field strength at the level $z = z_{\rm min}$. Assuming $\rho_{t=0}\equiv\rho(t=0,x,z)$ and $T_{t=0}\equiv T(t=0,x,z)$ as the initial normalized equilibrium density and temperature in a 2.5D domain, the gravitationally stratified density and temperature in the polytropic atmosphere are given by:
\begin{eqnarray}
  \rho_{t=0}&=&\left[ 1+\frac{(\gamma -1)}{\gamma H} \left(\frac{1}{10+z-z_{\rm min}} - \frac{1}{10}\right)\right]^{1/(\gamma -1)}, \label{equ:ide}\\
  T_{t=0}&=& \rho_{t=0}^{\gamma -1}.
  \label{equ:ite}
\end{eqnarray}
Here, $H =E_u F_r= 2k_B L_0 T_0/(m_pGM)$ denotes the normalized gravitational scale height,  with $k_B$ being Boltzmann constant, and $m_p$ represents the proton mass. The hydrostatic density scale height at the solar surface $H_0=(R^2_s/L_0)H=100 HL_0$ (assuming that $(z-z_{\rm min})\ll{R_s}$).

 We position a hot loop within the coronal arcade, with its edges following magnetic field lines, and one footpoint centered at ($x_0$,$z_0)$=(1, 1) with a radius (i.e., loop half width in the 2D model) to be $r_0=0.12$. The loop is characterized by having larger density and temperature inside than outside, with a peak density ratio $\chi_\rho=\rho_{\rm in}/\rho_{\rm ex}$ and a peak temperature ratio $\chi_T=T_{\rm in}/T_{\rm ex}$, where the subscript `in' refers to the interior of the loop and `ex' refers to its external surroundings). The temperature and density within the loop's cross section at the footpoint ($x_0$,$z_0$) are initialized with Gaussian profiles, given by
\begin{eqnarray}
\rho_{\rm ft}(r)&=&(\chi_\rho-1)\left(e^{-(2r/r_0)^2}-c_0\right)/(1-c_0)+1,\\
T_{\rm ft}(r)&=&(\chi_T-1)\left(e^{-(2r/r_0)^2}-c_0\right)/(1-c_0)+1.
\end{eqnarray} 
Here, $r=|x-x_0|\le{r_0}$ and the constant $c_0=e^{-4}$ is chosen to match the loop's boundary to the background corona. By considering $\rho_{t=0}(r)$ and $T_{t=0}(r)$ as the density and temperature at a point ($x$,$z$) along a field line that has the radial distance $r$ at the footpoint relative to ($x_0$,$z_0$), the loop's initial polytropic density and temperature satisfying the gravitational equilibrium are given by: 
\begin{eqnarray}
\rho_{t=0}(x,z,r)&=&\left[\rho_{\rm ft}^{\gamma-1}(r)+\frac{1}{c_r}\frac{(\gamma-1)}{\gamma{H}}\left(\frac1{10+z-z_{\rm min}}-\frac{1}{10}\right)\right]^{1/(\gamma-1)}, \label{equ:ild}\\
T_{t=0}(x,z,r)&=&c_r\rho_{t=0}^{\gamma-1}(x,z,r). \label{equ:ilt}
\end{eqnarray} 
Here, $c_r=T_{\rm ft}(r)/\rho_{\rm ft}^{\gamma-1}(r)$. This setting for the initial density and temperature distributions ensures that the density and temperature along the same field line within the loop have the approximately same contrast to the background corona. When $\chi_{\rho}=1$ and $\chi_T=1$, Equations \ref{equ:ild} and \ref{equ:ilt} reduce to Equations \ref{equ:ide} and \ref{equ:ite}, respectively. We also discuss the following two cases:

1) In the case where $\chi_{\rho}>1$ and $\chi_T=1$, the following relationships can be derived:
\begin{eqnarray}
  \rho_{t=0}^{\rm loop}&=&\rho_{\rm ft}\left[ 1+\frac{(\gamma -1)}{\gamma H} \left(\frac{1}{10+z-z_{\rm min}} - \frac{1}{10}\right)\right]^{1/(\gamma -1)}=\rho_{\rm ft}(r) \rho_{t=0}^{\rm bg}, \\
  T_{t=0}^{\rm loop}&=& 1+\frac{(\gamma -1)}{\gamma H} \left(\frac{1}{10+z-z_{\rm min}} - \frac{1}{10}\right)= T_{t=0}^{\rm bg},
\end{eqnarray}
where $\rho_{t=0}^{\rm loop}$ and $T_{t=0}^{\rm loop}$ represent the initial density and temperature distributions for the loop, while $\rho_{t=0}^{\rm bg}$ and $T_{t=0}^{\rm bg}$ represent those for the background (Equations \ref{equ:ide} and \ref{equ:ite}). This implies that the loop exhibits a density contrast of $\rho_{\rm ft}(r)$ relative to the background corona throughout the loop, while there is no temperature contrast. This loop model resembles the isothermal case \citep{selw09}.

2) In the case where $\chi_{\rho}=1$ and $\chi_T>1$, the following relationships can be derived:
\begin{eqnarray}
  \rho_{t=0}^{\rm loop}&=&\left[ 1+\frac{(\gamma -1)}{\gamma H T_{\rm ft}(r)} \left(\frac{1}{10+z-z_{\rm min}} - \frac{1}{10}\right)\right]^{1/(\gamma -1)}\\
  &\approx& 1 - \frac{0.01\Delta{z}}{\gamma H T_{\rm ft}(r)}, \\
  T_{t=0}^{\rm loop}&=&T_{\rm ft}(r) +\frac{(\gamma -1)}{\gamma H} \left(\frac{1}{10+z-z_{\rm min}} - \frac{1}{10}\right)\\
  &\approx& T_{\rm ft}(r) - \frac{(\gamma -1)}{\gamma H}(0.01\Delta{z}),
\end{eqnarray}
where $\Delta{z}=z-z_{\rm min}$. For $\Delta{z}=0$ at the footpoint and $\Delta{z}=0.72$ at the apex of the loop, we estimate the ratios of $\rho_{t=0}^{\rm loop}$ and $\rho_{t=0}^{\rm bg}$ to be 1 and 1.04, respectively. Moreover, we estimate $T_{t=0}^{\rm loop}/T_{t=0}^{\rm bg}\approx T_{\rm ft}(r)$ considering that $0.01\Delta{z}(\gamma-1)\ll{1}$ for $\gamma=1.05$. This suggests that the loop model displays a temperature contrast of approximately $T_{\rm ft}(r)$ compared to the background corona throughout the loop. Meanwhile, a very subtle density contrast ($\lesssim 4\%$) arises due to the larger scale height $T_{\rm ft}H$ in comparison to the background corona.

With the above setting, the loop's footpoint diameter measures 10 Mm at the coronal base (measured as the FWHM of the cross sectional density profile), and the loop length, $L$, is approximately 178 Mm, measured from a cut along the loop axis (see Figure~\ref{fig:init}(a)). Guided by AIA observations in \citet{wan15}, we take $T_0=7$ MK and $n_0=10^9$ cm$^{-3}$ with $\chi_T=1.5$ and $\chi_{\rho}$=1.5 in this study, resulting in the initial maximum temperature, $T_{\rm max}$, being 10.5 MK, and the maximum density, $\rho_{\rm max}$, being $1.5\times10^9$ cm$^{-3}$ in the hot loop model. Figure~\ref{fig:init} depicts the initial state of the coronal loop in the arcade magnetic field, encompassing density in Panel (a), temperature in Panel (b), plasma-$\beta$ represented as $\beta(x,z)=\beta_0(\rho T/B^2)$ in Panel (c), and local Alfv\'{e}n speed $V_A(x,z)=V_{A0}(B/\rho^{1/2})$ in Panel (d). We observe that the values of $\beta$ outside the loop increase with heights (from 0.0049 at the lower boundary to 0.69 at the upper boundary) while uniform along the $x$-direction in this arcade model. Within the loop, where $T\geq1$, the plasma-$\beta$ is slightly larger compared to the exterior but remains lower overall, ranging from 0.0049 to 0.05. This range aligns with typical coronal conditions as observed.

 The equations are solved in a 2.5D computational domain of size $(-2.4,\,2.4)\times (1,\,3.4)$, using normalized distance units and a uniform grid of $514\times514$ points. This resolution is verified to be adequate for the model through a convergence test using a higher resolution grid of $1026\times1026$. To run the 3D model in 2D mode, we keep 3 grid cells in the ($y$) direction perpendicular to the ($xz$) plane. The numerical method employed is the modified Lax-Wendroff method with a fourth-order stabilization term \citep[e.g.][]{OT02}.
 
It should be noted that the above initial state, including a dense and hot loop, is not an exact stable equilibrium, as the transverse pressure gradient in the loop is not balanced by magnetic pressure in the initial state. However, since the coronal loop is magnetically dominated, characterized by a low-$\beta$ condition, the departure from equilibrium is small. The relaxation process excites fast magnetoacoustic waves that propagate out of the domain after $t=10 \tau_A$. The relaxation also excites weak second harmonic slow-mode waves in the coronal loop (see Figure~\ref{fig:tdisv0.01}(a)), with a maximum velocity amplitude less than 2 km~s$^{-1}$ before we launch a flow pulse at $t=60 \tau_A$. This amplitude is about 200 times (10 times) smaller than that in Case 1 (Case 1B). Thus, its influence on our simulation results can be neglected.

\subsection{Boundary Conditions}
 The boundary conditions in the 2.5D computational domain are open on all external planes except for the bottom boundary located at the coronal base $z=z_{\rm min}$. At this lower boundary, the perturbed magnetic fields, temperature, and velocities are set to be zero, while the density is extrapolated from the values obtained at the interior points. To excite the slow magnetoacoustic waves, a velocity pulse is impulsively injected at the right footpoint of the coronal loop model (centered at $x_0$=1 in the area $r=|x-x_0|\leq 2w_v$=0.06, i.e., within the loop's cross section at $z=z_{\rm min}$). A similar method has been applied in previous studies \citep{wan13, prov18, ofm22}. The injected velocity is along the magnetic field direction and given by:
\begin{equation} \mathbf{V} = V_0(x,z_{\rm min},t)  \frac{\mathbf{B}}{|B|},
\end{equation}
where the dimensionless velocity $V_0$ is given by:
\begin{equation}
V_0(x,z=z_{\rm min},t)  = A_v(t) \exp\left[ -\left( \frac{r}{w_v} \right)^4 \right].
\end{equation}
Here, $A_v(t)$ sets the time profile of a single pulse as:
\begin{eqnarray}
A_v(t) =  \left\{
\begin{array}{ll}
A_0 \left[\frac{1}{2}\left(1 - \cos \frac{2 \pi (t-t_1)}{\Delta{t}}\right) \right]& ~~{\rm for~} t_1 \leq t \leq t_2,\\
0 & ~~{\rm otherwise},
\end{array}
\right.  
%\nonumber
\end{eqnarray}
with the time interval between $t_1=60\tau_A$ and $t_2 = 90\tau_A$, and the magnitude $A_0$=0.1, approximately 2 times $C_{s0}$. The pulse duration is chosen to be $\Delta{t}=30\tau_A$, which is much shorter than the anticipated slow-mode wave period in the loop, as guided by observations \citep{wan05, wan18}.

A similar condition is applied for the dimensionless temperature perturbation $\Delta{T}$:
\begin{equation} 
\Delta{T}(x,z_{\rm min},t) =A_T(t) \exp\left[ -\left( \frac{r}{w_v} \right)^4 \right] ~~(r \leq 2w_v),
\end{equation}
where the function $A_T(t)$ is based on the relationship, $\frac{\Delta{T}}{T_0}=(\gamma-1)\frac{V}{C_s}$, derived from linearized ideal MHD theory:
\begin{eqnarray}
    A_T(t)= \left\{
\begin{array}{ll}
    \frac{(\gamma-1)V_{A0}}{C_{s0}}A_v(t) &  ~~{\rm for~} t_1 \leq t \leq t_2 ,\\
  0 & ~~{\rm otherwise}.   
\end{array}   
\right.  
\end{eqnarray}
Based on the values of $V_{A0}$ and $C_{s0}$ provided in Table~\ref{tab:para}, it is estimated that $A_T(t)\approx A_v(t)$ in normalized units. This implies that the maximum amplitude in the temperature pulse is about 0.1$T_0$ for the case of $A_0$=0.1. Outside the pulse source region ($r>2 w_v$), we set the velocity and temperature perturbations to zero at $z=z_{\rm min}$.

\section{Results}
\label{sct:rst}
Motivated by observational analyses of slow-mode waves in a hot flaring loop in \citet{wan15} and previous 1D MHD modeling \citep{wan18,wan19} and 3D MHD modeling \citep{ofm22}, our main focus is on comparing two types of models using 2.5D MHD simulations of an arcade AR initiated with a hot ($\chi_T=1.5$) and dense ($\chi_\rho=1.5$) coronal loop. Model 1 incorporates the classical thermal conduction coefficient ($\kappa_\|=\kappa_c$) and the classical compressive viscosity coefficient ($\eta_0=\eta_c$). On the other hand, Model 2 maintains $\kappa_\|=\kappa_c$ but increases $\eta_0$ to 10 times the value of $\eta_c$. The purpose of this comparison is to quantitatively examine the effects of anomalous viscosity on the excitation and evolution of slow-mode waves. Additionally, we conduct control numerical experiments by comparing these two models under different sets of parameters, as outlined in Table~\ref{tab:cases}. 

In Case A, we set $\kappa_c$ to 0 to establish a reference condition without thermal conduction. This allows us to analyze the influence of thermal conduction on the wave behavior. In Case B, we reduce the initial flow amplitude by a factor of 10 ($A_0=0.01$) to investigate the effects of nonlinearity on wave evolution. Case C examines the impact of density structure on the wave properties by reducing a density contrast of $\chi_\rho$ to 1. Conversely, in Case D, we explore the effect of thermal structure by reducing a temperature contrast of $\chi_T$ to 1. Case E investigates the impact of hotter loop structure on wave trapping by increasing a temperature contrast of $\chi_T$ to 2. Case F excludes the viscous term in the energy equation to assess its contribution to wave modeling. 

\begin{figure}
\centerline{\includegraphics[width=1.0\textwidth,clip=]{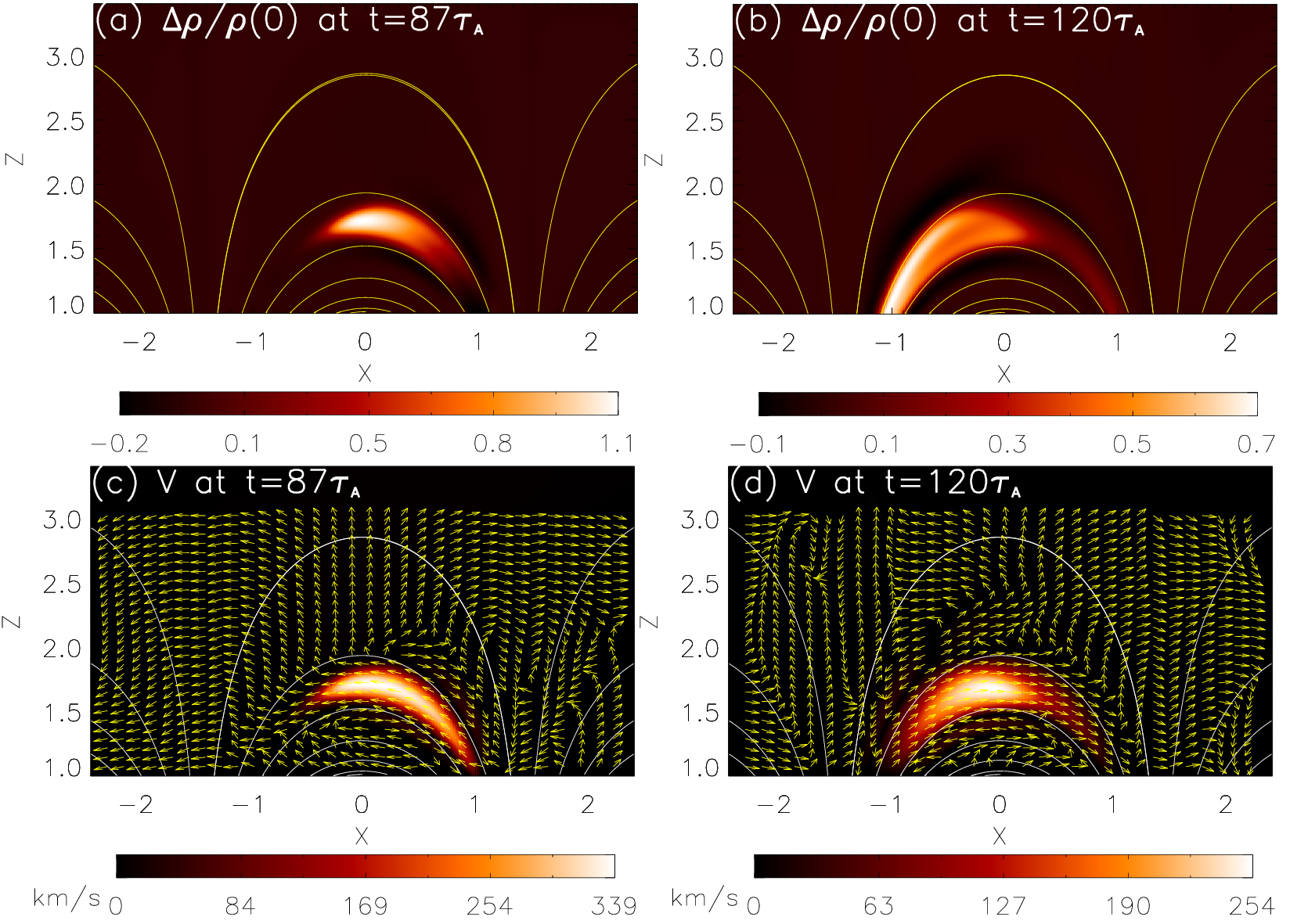}}
\caption{Snapshots of relative perturbed density ($\Delta{\rho}/\rho(0)=(\rho(t)-\rho(0))/\rho(0)$) and velocity in the $xz$-plane for Model 1 with classical thermal conduction and compressive viscosity coefficients. (a) and (b): density distributions at $t=87\tau_A$ and $120\tau_A$. (c) and (d): Corresponding velocity distributions at $t=87\tau_A$ and $120\tau_A$. The intensity scale represents the magnitude of velocities, and the arrows indicate their directions. The solid lines in each panel depict the magnetic field lines. An animation for this figure is available in the online journal.}
\label{fig:nvr1}
\end{figure}

\begin{figure}
\centerline{\includegraphics[width=1.0\textwidth,clip=]{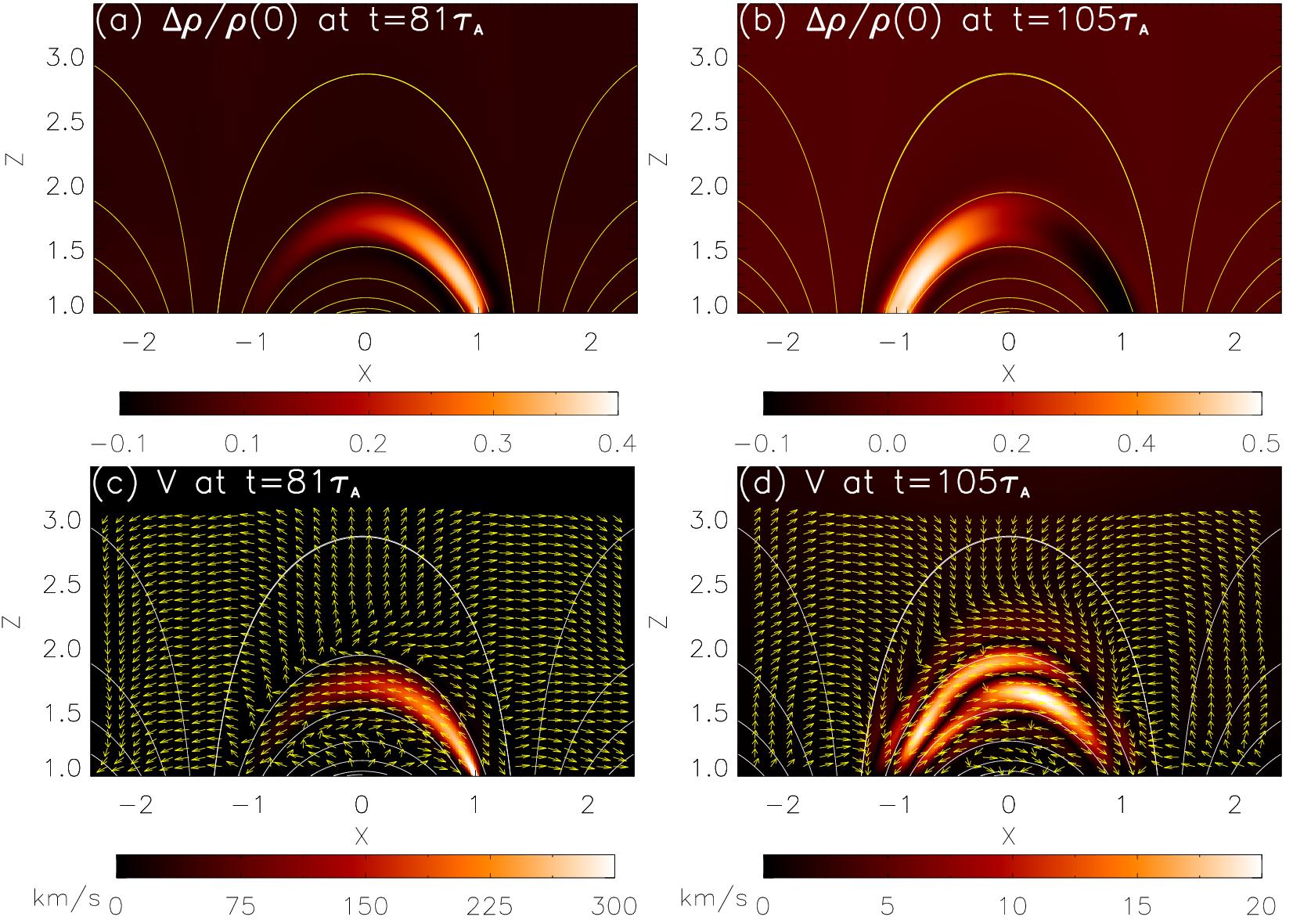}}
\caption{Same as Figure~\ref{fig:nvr1} but for Model 2 with classical thermal conduction and a 10-times enhanced compressive viscosity. (a) and (b): Density distributions at $t=81\tau_A$ and $105\tau_A$. (c) and (d): Corresponding velocity distributions at $t=81\tau_A$ and $105\tau_A$. The intensity scale shows the magnitude of velocities, and the arrows show their directions. An animation for this figure is available in the online journal. }
\label{fig:nvr2}
\end{figure}

\subsection{Effect of Enhanced Compressive Viscosity}
\label{sct:ecv}
We compare the simulation results of waves excited in a hot and dense coronal loop by a velocity pulse at the right footpoint (\textit{flare source}) between Cases 1 and 2, as shown in Figures 2-6. Figure~\ref{fig:nvr1} presents snapshots of perturbed density and velocity in the ($x$,$z$)-plane for Case 1 at two times: $t=87\tau_A$, when the initial wave front reaches the loop apex (Panel (a)), and $t=120\tau_A$, when the wave front reflected from the left footpoint (\textit{remote footpoint}) reaches the loop apex (Panel (b)). Similarly, Figure~\ref{fig:nvr2} displays snapshots of perturbed density and velocity for Case 2 at $t=81\tau_A$, when the initial density perturbation propagates towards the left footpoint (Panel (a)), and $t=105\tau_A$, when the wave is reflected from the left footpoint (Panel (b)). There are noticeable differences in wave features between Case 2 and Case 1. In Case 2, where enhanced viscosity is considered, the initial density perturbation is greatly elongated along the loop, without a discernible wave front as observed in Case 1. Another evident difference is that in Case 1, both the initial and reflected perturbations of density and velocity are in-phase. However, in Case 2, only the initial perturbations of density and velocity are in-phase. When the density perturbation reaches its maximum near the remote footpoint, the velocity perturbation becomes much weaker, with inward and outward propagating waves mixed (refer to Figure~\ref{fig:nvr2}(d)). 

The online animations corresponding to Figures~\ref{fig:nvr1} and \ref{fig:nvr2} are available. In these animations, both Cases 1 and 2 exhibit similar ``sloshing motions", where a density enhancement appears to oscillate back and forth along the loop between the two footpoints. However, their difference can be clearly distinguished through time-distance plots. It is known that for a fundamental standing slow-mode wave in a loop, density perturbation is characterized with a node structure at the loop apex and the antiphase oscillations between the two legs, while velocity perturbations along the loop are in phase \citep[see][]{yuan15}. In contrast, the reflected propagating wave exhibits a ``zigzag" pattern in the time-distance plot \citep[see][]{wan18,wan21}. 

\begin{figure}
\centerline{\includegraphics[width=1.0\textwidth,clip=]{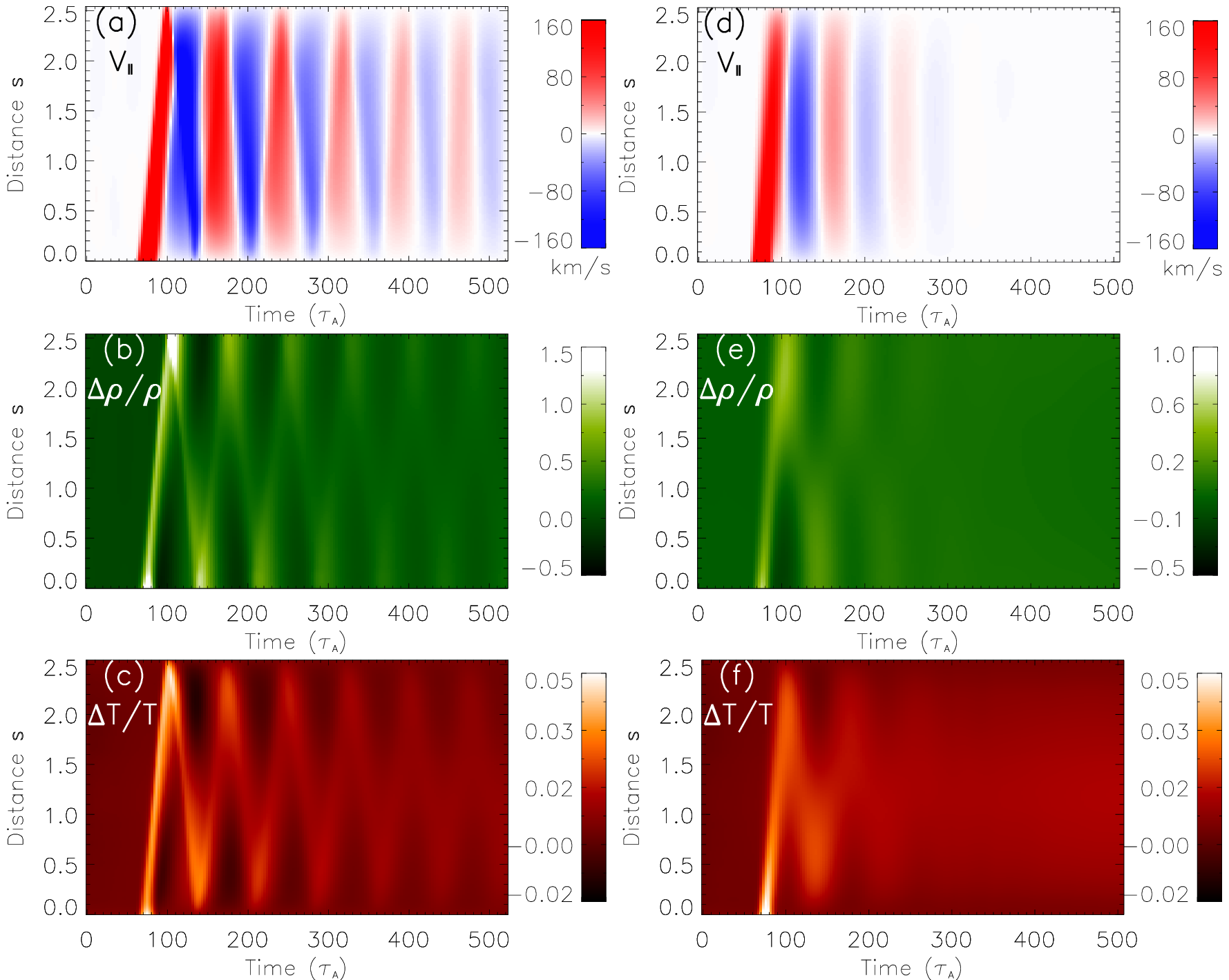}}
\caption{\textit{Left panels:} Time distance maps of (a) velocity parallel to the magnetic field ($V_\|$), (b) relative perturbed density ($\Delta\rho/\rho(0)$), and (c) relative perturbed temperature ($\Delta{T}/T(0)$) for a slice along the loop in Model 1. Here, $\rho(0)$ and $T(0)$ represent the initial density and temperature distributions along the loop, respectively. \textit{Right panels:} Same as the left panels but for Model 2. The distance $s$ (in units of $a$=70 Mm) is measured from the right footpoint of the loop.}
\label{fig:tdis}
\end{figure}

Figure~\ref{fig:tdis} compares the time-distance maps of the magnetic field-parallel velocity component (Top Panels), relative perturbed density (Middle Panels), and relative perturbed temperature (Bottom Panels) for Cases 1 and 2. Based on the zigzag feature observed in the density perturbation (see Panel (b)), we can qualitatively identify that Case 1 exhibits a reflected propagating wave with a weaker damping rate during the entire simulated period. Whereas the presence of the two-leg anti-phase oscillations in Case 2 indicates the rapid formation of a standing mode wave with a strong damping rate (comparable to observations) after a single reflection of the initial pulse (see Panel (e)). This result confirms the conclusions drawn from the 1D and 3D MHD simulations with similar model parameters \citep{wan18,wan19,ofm22}. 

Additionally, the formation of a standing wave can be tentatively identified from the (alternate red/blue) parallel bars pattern observed in the time-distance map of velocity, indicating in-phase oscillation along the loop. Figure~\ref{fig:tdis}(a) suggests the initiation of a standing wave when $t>410\tau_A$. However, the density map still displays a zigzag pattern (more visible with increased contrast), implying that the standing wave has not fully established. Similarly, Figure~\ref{fig:tdis}(d) suggests the development of a standing wave when $t>110\tau_A$, aligning with the transition time identified from the density pattern.

\begin{figure}
\centerline{\includegraphics[width=1.0\textwidth,clip=]{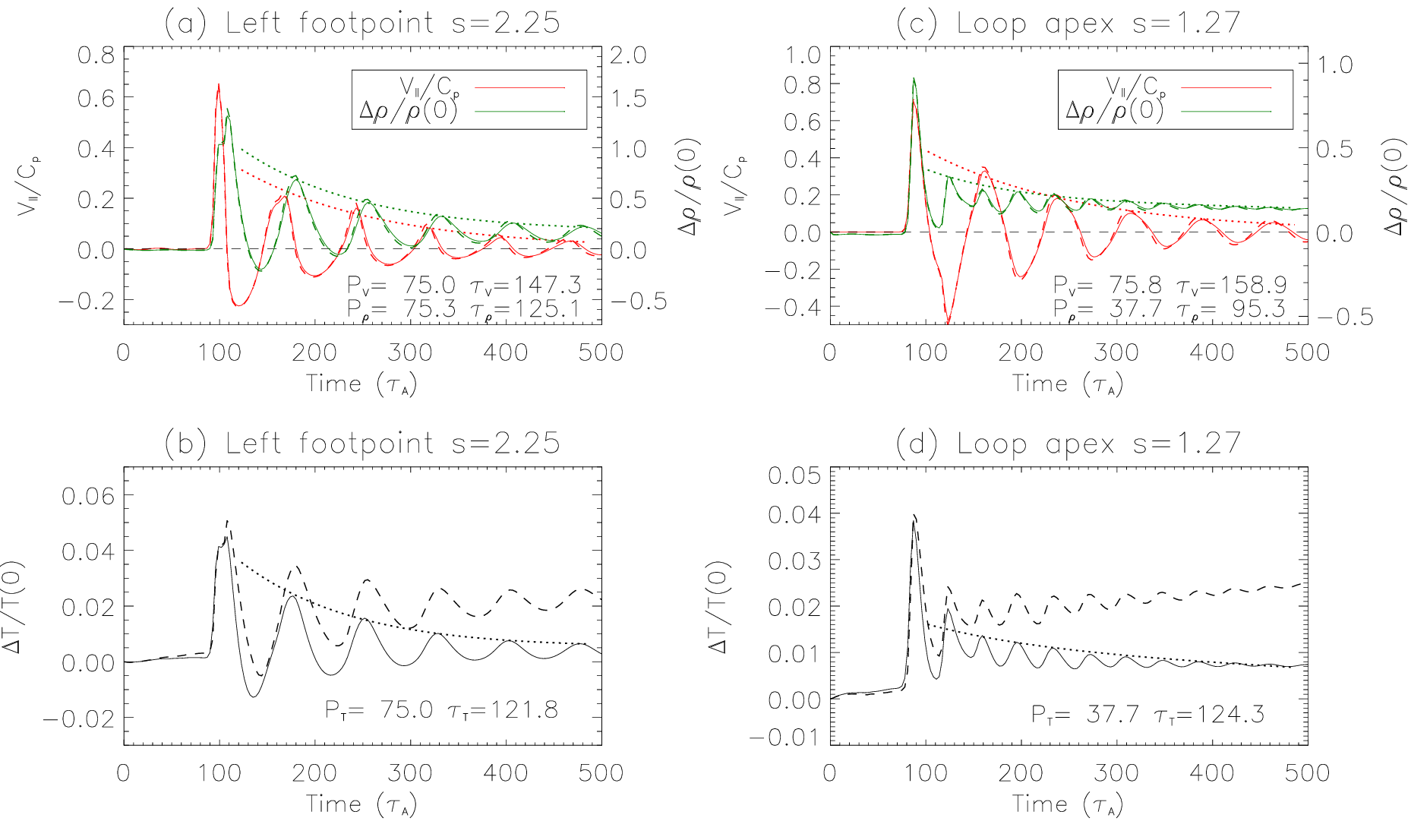}}
\caption{Temporal evolution of (a) the velocity component parallel to the magnetic field ($V_{\|}$ in units of the local sound speed $C_p$ at the initial temperature $T(0)$) and the perturbed density ($\Delta\rho/\rho(0)=\frac{\rho(t)-\rho(0)}{\rho(0)}$), and (b) the perturbed temperature ($\Delta{T}/T(0)=\frac{T(t)-T(0)}{T(0)}$) at the location $s$=2.25 near the left footpoint of the loop in Model 1. The dotted lines indicate the exponential decay time fit. The measured oscillation periods and decay times are marked on the plots. The dashed lines have the same meaning as the solid lines but for the case without thermal conduction. Panels (c) and (d) show the same as (a) and (b) but at the location $s$=1.27 near the loop apex.  Note that in (a) and (c), the dashed curves are closely overlaid with the solid ones.} 
\label{fig:prf1}
\end{figure}

\begin{figure}
\centerline{\includegraphics[width=1.0\textwidth,clip=]{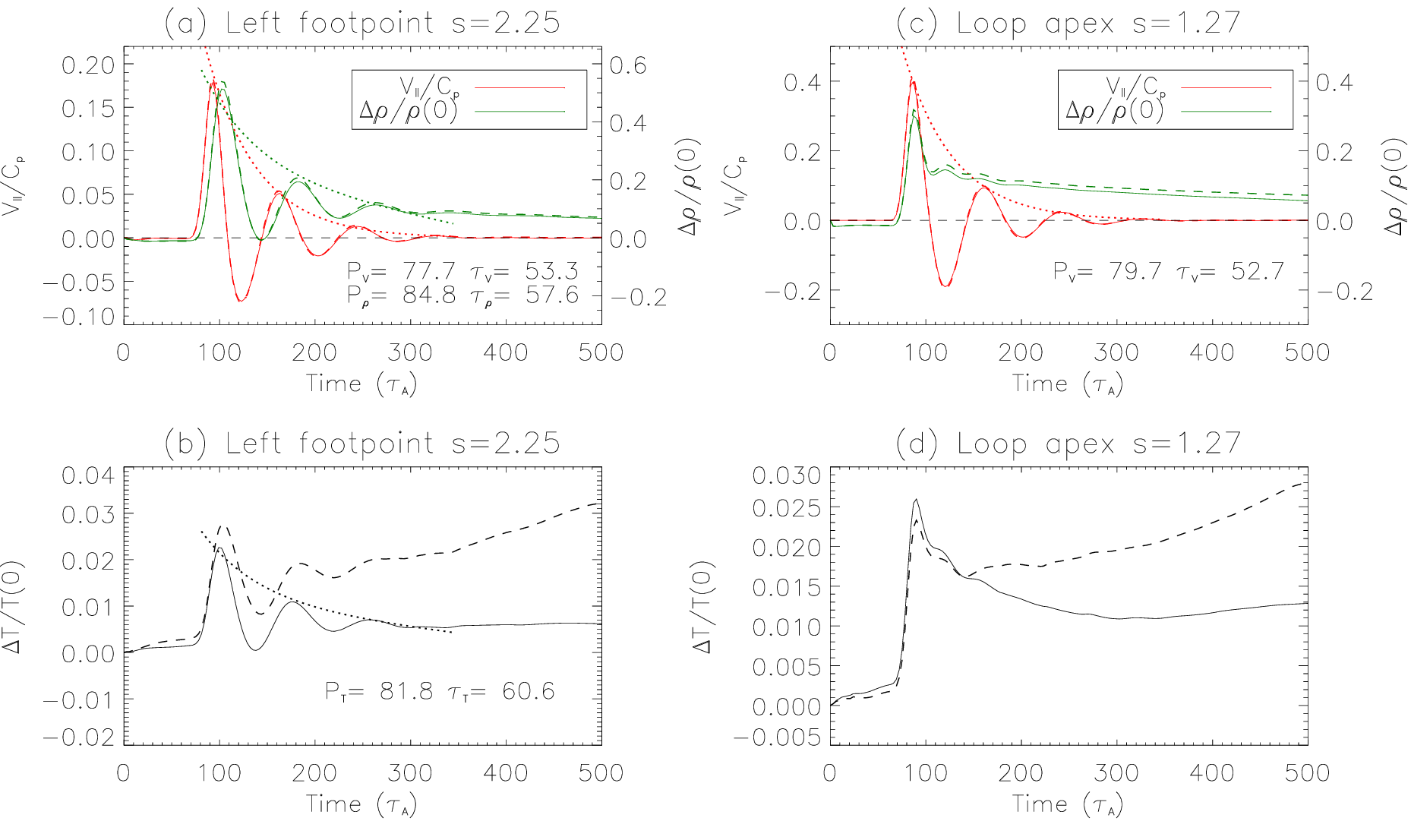}}
\caption{Temporal evolution of (a) the velocity component parallel to the magnetic field ($V_{\|}/C_p$) and the perturbed density ($\Delta\rho/\rho(0)$), and (b) the perturbed temperature ($\Delta{T}/T(0)$) at the location $s$=2.25 near the left footpoint for Model 2. The dotted lines indicate the exponential decay time fit. The measured oscillation periods and decay times are marked on the plots. The dashed lines have the same meaning as the solid lines but for the case without thermal conduction. Panels (c) and (d) show the same as (a) and (b) but at the location $s$=1.27 near the loop apex.  Note that in (a) and (c), the dashed curves are closely overlaid with the solid ones.} 
\label{fig:prf2}
\end{figure}

We can quantitatively determine whether or when the initially generated wave pulse transitions from propagating into standing mode based on the phase relationship between the field-parallel velocity and density oscillations at the loop leg or near the apex \citep[e.g.,][]{selw05,selw07}. In the standing mode, there is a phase shift of 90$^{\circ}$ while the propagating wave exhibits a phase shift of 0$^{\circ}$ when not considering the sign of velocity.  It should be noted that for a reflected propagating wave, the phase shift measured from a location near the footpoints is also close to 90$^{\circ}$ due to the superposition effect of inward and outward parts of the wave pulse during its reflection (e.g., the case shown in Figure~\ref{fig:prf1}(a)). Thus, based on the in-phase relationship between velocity and density oscillations as shown in Figure~\ref{fig:prf1}(c) for the apex, we determine that the reflected wave in Case 1 has not yet transitioned into a standing mode throughout the simulation time. In contrast, in Case 2, the phase shift reaches approximately 90$^{\circ}$ after $t=140\tau_A$, measured at the loop leg ($s=3/4 L$), indicating the establishment of a standing mode. The result is not shown but is similar to Figure~\ref{fig:prf2}(a). The amplitudes of density perturbation at the apex reduce to nearly zero after $t=140\tau_A$ (Figure~\ref{fig:prf2}(c)). This indicates the formation of a null point at the apex, further confirming this transition.

To measure the physical parameters of simulated waves and compare their evolution between Cases 1 and 2, we plot the time profiles of velocity, perturbed density, and perturbed temperature at two locations along the loop. These locations are $s$=2.25 near the remote footpoint (Left Panels) and $s$=1.27 near the loop apex (Right Panels) in Figures~\ref{fig:prf1} and \ref{fig:prf2}. We estimate the wave period ($P$) by averaging the time intervals between successive peaks in the profile of each variable, and we determine the damping time ($\tau$) by fitting the wave peaks to an exponentially damped function ($f(t)=A_0+A_1{t}+A_2{\rm exp}(-t/\tau)$). The measured values of $P$ and $\tau$ are indicated on the plots and listed in Table~\ref{tab:ptd}. It should be noted that in Case 1, we have excluded the first peak when fitting the time profile of perturbations to avoid the influence of the initial flow on  wave damping measurements.

At the footpoint in Case 1, we observe that the measured periods for velocity, density, and temperature oscillations are the same ($P=75\tau_A$). However, at the loop apex, the periods for density and temperature oscillations are half of the velocity period. This is because the loop apex experiences two perturbations from the inward and reflected waves during each period of velocity oscillation. By using the loop length and wave period, we can estimate the wave phase speed as $V_p=2L/P$, resulting in $V_p$=470 km~s$^{-1}$. This value roughly agrees with the sound speed ($C_p=426$ km~s$^{-1}$ with $\gamma$=1.05) for the maximum temperature in the loop of $T_m=\chi_T\,T_0$=10.5 MK, confirming that the excited waves are slow-mode waves.

Damping times measured between the velocity and density (temperature) oscillations are slightly different at both the footpoint and the apex. At the footpoint, the damping time to oscillation period ratio (often referred to as the oscillation quality factor, $Q=\tau/P$) is 2.0 for $V_\|$, while it is 1.7 (1.6) for $\Delta{\rho}$ ($\Delta{T}$). At the apex, the $Q$-factor for $V_\|$ is 2.1, whereas $Q$=2.5 (3.3) for $\Delta{\rho}$ ($\Delta{T}$). The higher $Q$-factor for density and temperature perturbations at the apex is primarily due to the shorter periods of oscillation for these variables.

In Case 2, we find that the average oscillation period for velocity, density, and temperature perturbations at the footpoint is approximately 82$\tau_A$, and the average oscillation $Q$-factor is around 0.7. These values closely resemble those observed for velocity oscillation at the apex. It is noteworthy that the amplitudes of density and temperature oscillations at the loop apex are considerably smaller compared to those of velocity. This characteristic is consistent with a standing slow wave, where there is an anti-node in velocity and a node in density and temperature perturbations at the loop apex. We also observe that the measured oscillation periods in Case 2 are slightly longer than those in Case 1. This difference may be attributed to the increased viscosity in Case 2 compared to Case 1, as discussed in \citet{wan19}.

\subsection{Effect of Thermal Conduction}
\label{sct:etc}
We will now analyze the effects of thermal conduction on the evolution of the excited waves by comparing the above results with Case A, where the thermal conduction term is disabled ($\kappa_\|$=0) in simulations of Cases 1 and 2. The dashed lines in Figures~\ref{fig:prf1} and \ref{fig:prf2} illustrate the evolution of velocity, density, and temperature perturbations at the footpoint and apex of the loop in these modified cases. We observe that, for both Cases 1A and 2A, the impact of thermal conduction on velocity and density perturbations is negligible. This outcome is expected since we have assumed a nearly isothermal condition by employing a polytropic index of $\gamma=1.05$ in these simulations. Consequently, the amplitudes of temperature perturbations shown in Figures~\ref{fig:prf1} and \ref{fig:prf2} are significantly smaller (by an order of magnitude) compared to those of velocity and density perturbations. Therefore, any changes in the evolution of thermal energy (related to $T$) have minimal effect on the evolution of kinetic energy (related to $V$). 

We also note a prominent difference in Case 1A (without thermal conduction) compared to Case 1, which is the gradual increase in the background trend of temperature perturbations (Panels (b) and (d) of Figure~\ref{fig:prf1}). This behavior can be attributed to viscous heating, as its effect is counteracted by the cooling from thermal conduction when included in Case 1. In Case 2A, the background temperature at both the footpoint and apex increases much more rapidly compared to Case 1A (Panels (b) and (d) of Figure~\ref{fig:prf2}). This discrepancy can be explained by the fact that the viscous heating rate in Case 2A is 10 times higher than in Case 1A, owing to the viscosity coefficient $\eta_0$ being 10 times larger \citep[see][]{ofm02, wan19}. Although the inclusion of thermal conduction leads to a noticeable change in the loop's background temperature, its effect on the wave damping rate is weak due to our assumption of $\gamma$=1.05. The measurements indicate that the $Q$-factor for temperature perturbations is approximately 10$-$20\%  smaller in Case 1 compared to Case 1A, and a similar effect is observed in Case 2. Finally, we observe from Figures~\ref{fig:prf1}(b) and \ref{fig:prf2}(b) that thermal conduction causes a slight phase shift (peaking earlier by a few $\tau_A$) in temperature perturbations for Cases 1 and 2 compared to the temperature perturbations in Case A or the density perturbations. This behavior becomes more pronounced under the same physical conditions when $\gamma$=5/3 \citep{wan19}.

\subsection{Effect of Viscous Term in Energy Equation}
\label{sct:evt}
 We investigate the impact of energy transfer resulting from compressive viscosity on the behavior of simulated waves in Cases 1 and 2 by deactivating the viscous term in the energy equation (Case F). Comparing Case 1 with Case 1F, we observe no significant difference in the evolution of velocity, density, and temperature perturbations (not shown). This suggests that the dissipation of wave energy through viscous heating, compared to other mechanisms like thermal conduction and wave leakage, is negligible when using the classical viscosity coefficient. However, this effect becomes noteworthy in Case 2, where the viscosity is substantially enhanced, as shown in the following analysis. 

\begin{figure}
\centerline{\includegraphics[width=1.0\textwidth,clip=]{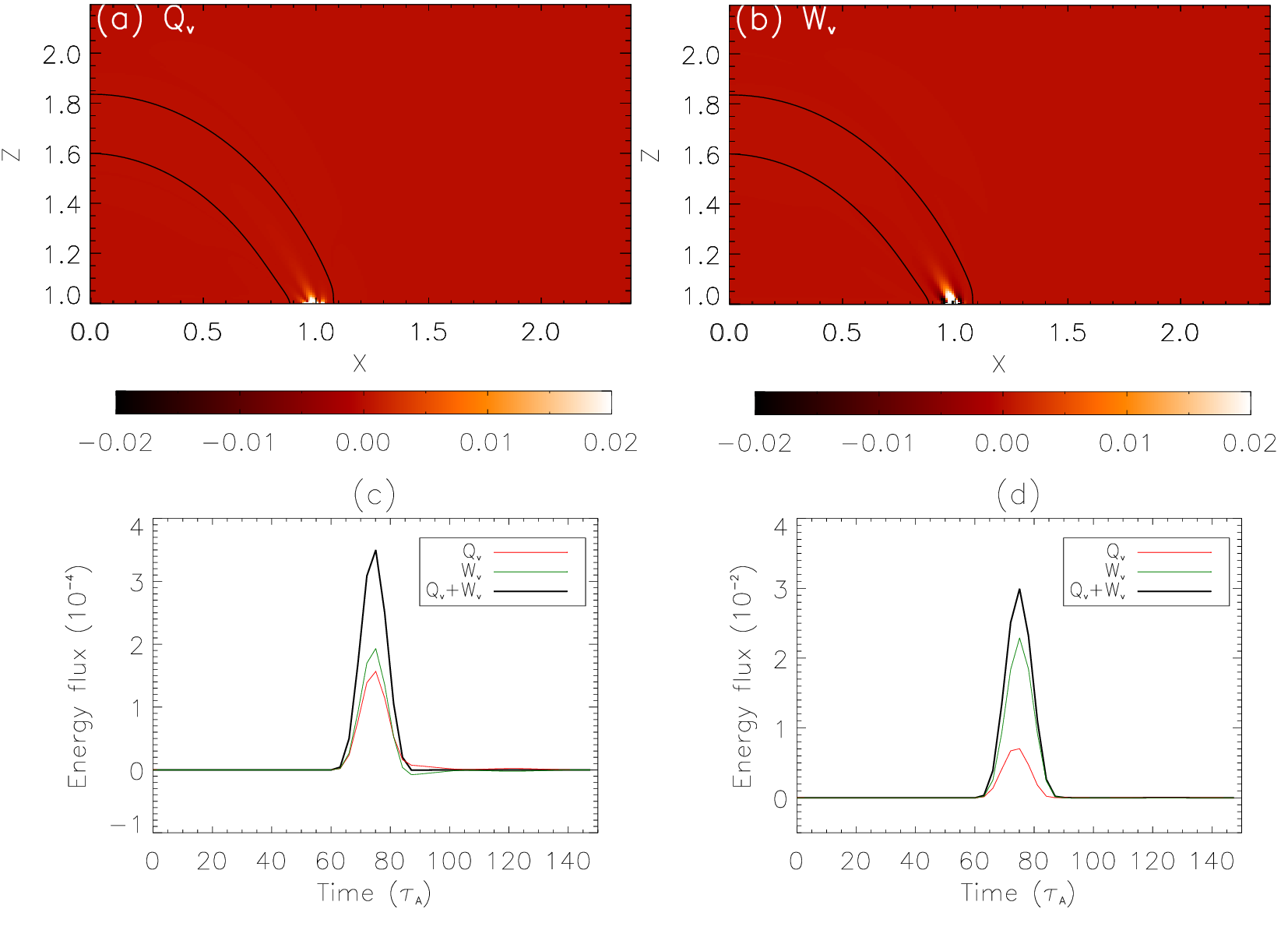}}
\caption{Contribution of the viscous term to energy transfer for Case 2. Distributions of (a) viscous heating rate ($Q_v$) and (b) energy transfer rate by viscous force ($W_v$) in the bottom-right quarter of the simulation domain at $t=75\tau_A$. The black contours indicate the loop region with $\rho\ge1.1$. Panels (c) and (d) show the time evolution of $Q_v$ (red  line), $W_v$ (green line), and their sum $Q_v+W_v$ (black thick line), averaged over the entire loop region with $\rho\ge1.1$ and  for a point ($x$,$z$)=(0.98,1.05) near the right footpoint, respectively. }
\label{fig:qwm}
\end{figure}

Figures~\ref{fig:qwm}(a) and (b) display the distributions of the viscous heating rate ($Q_v$) and the energy transfer rate ($W_v$) by viscous force  in the energy equation at the peak time of the footpoint driving flow, $t=75\tau_A$. Since their contributions mainly concentrate near the right footpoint where the initial flow is injected, we only display the bottom-right part of the simulation domain. Temporal evolutions of $Q_v$, $W_v$, and their sum, averaged over the entire loop region with a density $\rho\ge1.1$, and for a small region near the center of the right footpoint, are shown in Figures~\ref{fig:qwm}(c) and (d), respectively. We observe that $Q_v$ and $W_v$ exhibit comparable amplitudes and peak simultaneously, aligning with the flow driver ($V_0(x,z=z_{\rm min},t)$) at the bottom boundary. 

\begin{figure}
\centerline{\includegraphics[width=1.0\textwidth,clip=]{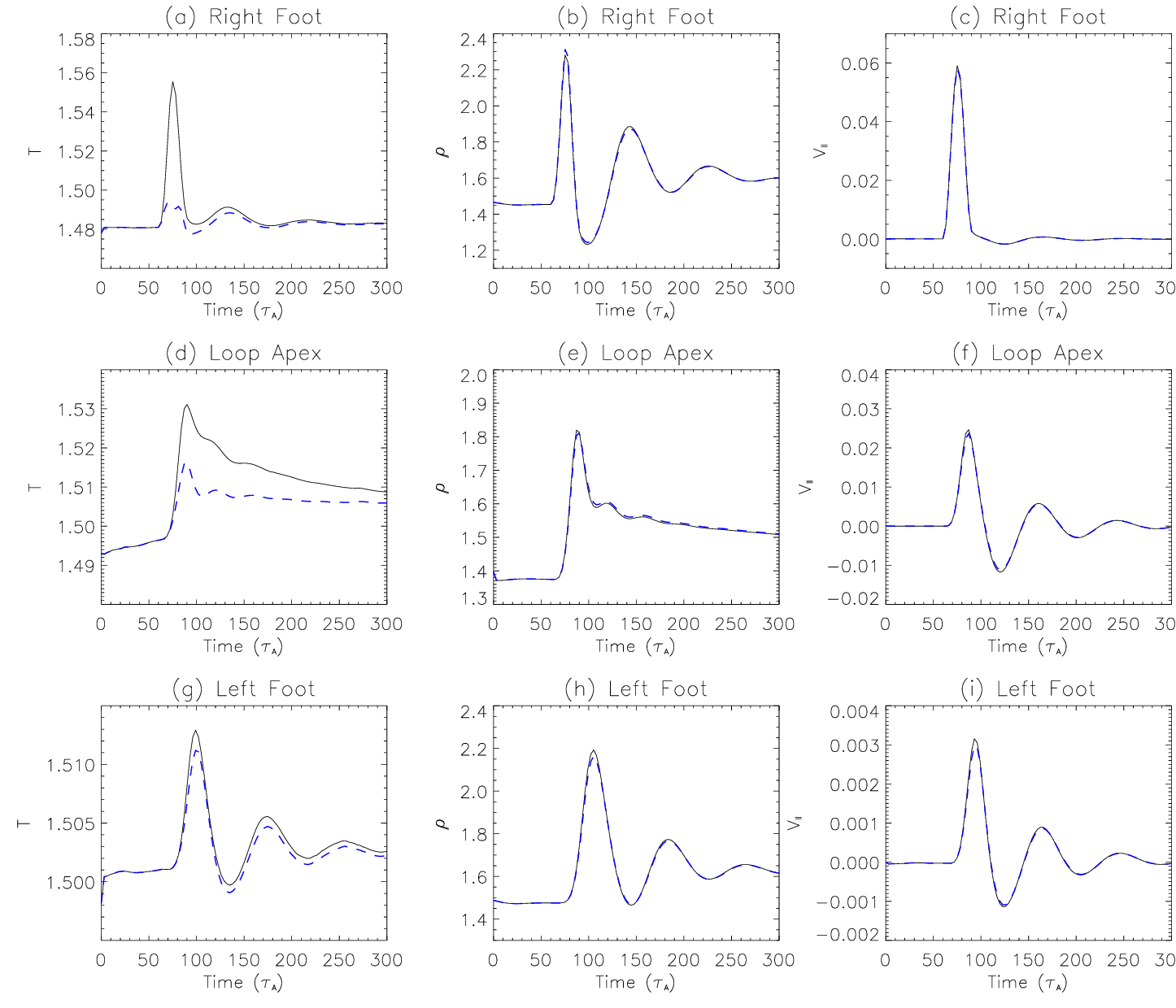}}
\caption{Comparison between Case 2 (with the viscous term) and Case 2F (without the viscous term) in the energy equation. Panels (a)-(c): Time profiles of temperature ($T$), density ($\rho$), and velocity parallel to the magnetic field ($V_\|$) at a position near the right footpoint with the coordinate ($x$,$z$)=(0.97,1.1). Panels (d)-(f): Similar to panels (a)-(c) but at a position near the loop apex with coordinates ($x$,$z$)=(0.0,1.7). Panels (g)-(i): Similar to panels (a)-(c) but at a position near the left footpoint with coordinates ($x$,$z$)=(-0.96,1.1). Solid lines represent Case 2, while dashed lines represent Case 2F. Note that the dashed curve for panels in the middle and right columns is closely overlaid with the solid curve.}
\label{fig:qwf}
\end{figure}

From Panel (b) of Figure~\ref{fig:qwm}, we notice that $W_v$ is positive in the central region of the loop, leading to an increase in mechanical energy of the plasma, while it is negative towards the loop edge, playing an opposite role. However, the average of $W_v$ over the entire loop is positive, indicating that the net effect of energy transfer by the viscous force is to add mechanical energy into the loop during the flow injection phase (Figure~\ref{fig:qwm}(c)). In contrast, the viscous heating in the energy equation consistently increases the internal energy of the plasma and is associated with an increase in temperature.

In Figure~\ref{fig:qwf}, we compare the evolution of velocity, density, and temperature at different locations within the loop. We observe that in Case 2, the temperature variations have a much larger amplitude compared to Case 2F, especially at the right footpoint and the apex of the loop, indicating a strong effect of viscous heating caused due to the enhanced viscosity. However, the differences in velocity and density variations between Case 2 and Case 2F are negligible. This result can be attributed to the fact that the increase of mechanical energy (primarily kinetic energy) within the loop, resulting from the work done by the viscous force, is much smaller than the kinetic energy directly injected from the loop's footpoint. 

It is worth noting that the damping of wave amplitudes in Cases 2 and 2F is predominantly related to the dissipation effect of the viscous force in the momentum equation, resulting in a reduction in the wave's kinetic energy. The limited influence of neglecting the viscous heating term in the energy equation can be possibly attributed to the assumption of $\gamma$=1.05, which implies an almost isothermal condition. In such a scenario, temperature fluctuations exhibit significantly smaller amplitudes compared to those in velocity and density. Consequently, pressure variations remain largely unaffected by viscous heating (which mainly influences temperature). As a result, the velocity evolution remains relatively unaltered since it is primarily governed by the pressure gradients for the slow waves.

\subsection{Effect of Nonlinearity}
\label{sct:enl}
1D MHD simulations have shown that slow waves with large amplitudes can lead to generation of higher harmonics and wave-front steepening due to nonlinearity \cite[e.g.,][]{ofm02,verw08}, while 2D and 3D MHD simulations suggest the additional effect of nonlinearity such as mode coupling and wave leakage \cite[e.g.,][]{dem04b, selw07, ofm12, ofm22}. These nonlinear effects play important roles in affecting the wave excitation and damping. 

We investigate the nonlinear effect regarding the wave-front evolution by comparing Cases 1 and 2, excited by the flow injection with the maximum amplitude $A_0=0.1$ (approximately 2 times $C_{s0}$), with a control numerical experiment with an amplitude 10 times smaller (Case B) in this section, while the nonlinear effects regarding mode coupling and wave leakage will be explored in Section~\ref{sct:emc}. 

\begin{figure}
\centerline{\includegraphics[width=1.0\textwidth,clip=]{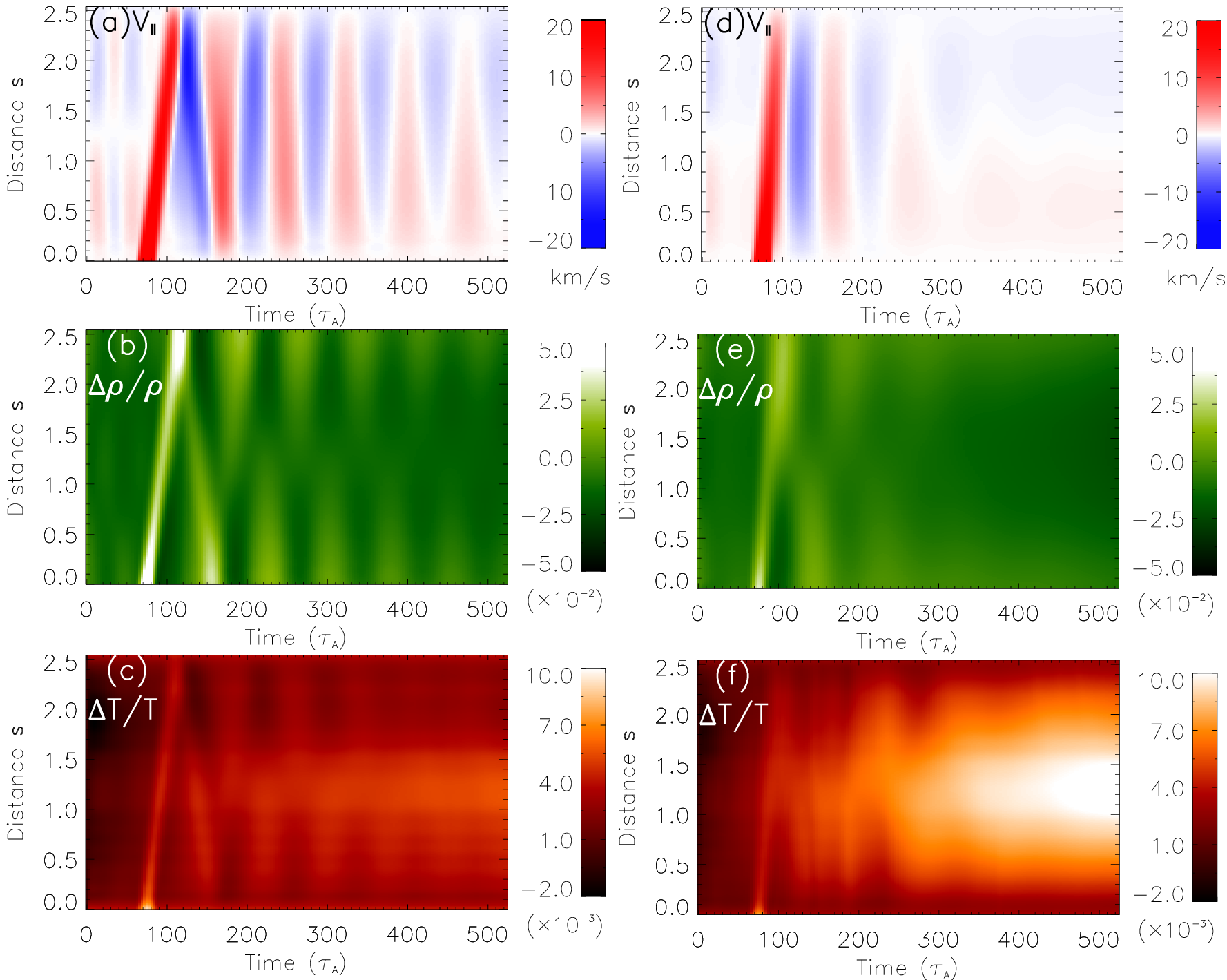}}
\caption{ 
Simulation results for Case B with an initial flow amplitude $A_0=0.01$. \textit{Left panels} display time-distance maps of (a) field-parallel velocity, (b) relative density perturbation, and (c) relative temperature perturbation for a slice along the loop in Case 1B. \textit{Right panels} present equivalent measurements, but for Case 2B.}
\label{fig:tdisv0.01}
\end{figure}

\begin{figure}
\centerline{\includegraphics[width=1.0\textwidth,clip=]{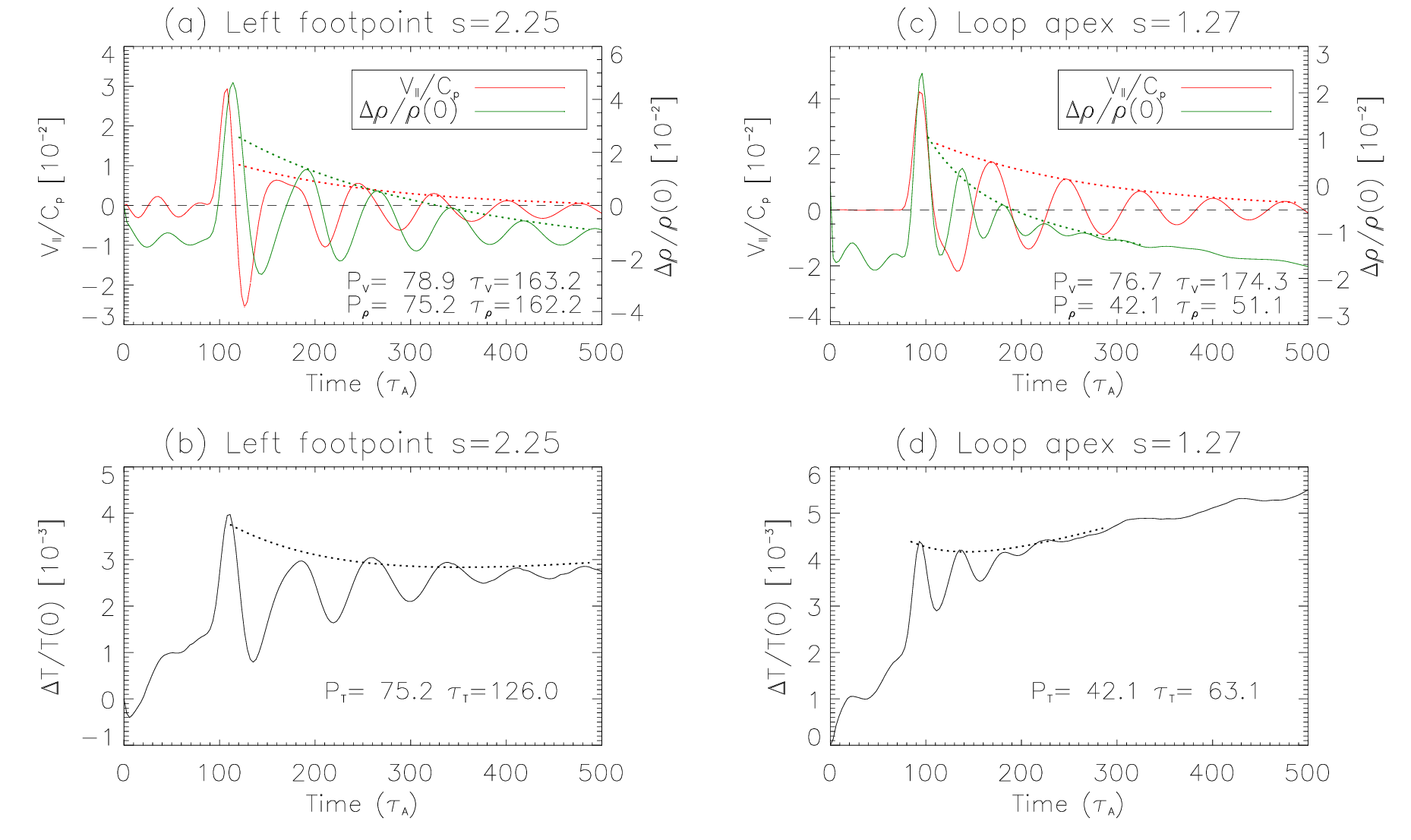}}
\caption{ 
 Simulation results for Case 1B with $A_0=0.01$. Temporal evolution of (a) the field-parallel velocity component relative to the initial local sound speed $C_p$ and the relative density perturbation, and (b) the relative temperature perturbation at the location $s$=2.25 near the left footpoint of the loop. The dotted lines indicate the exponential decay time fit. Panels (c) and (d) display the same parameters as in panels (a) and (b), respectively, but at the location $s$=1.27 near the loop apex.}
\label{fig:prf1v0.01}
\end{figure}

Figure~\ref{fig:tdisv0.01} shows the evolution of axis-parallel velocity, relative perturbed density, and temperature along the loop for Cases 1B and 2B. In Case 1B, density perturbations along each leg synchronize rapidly, exhibiting anti-phase oscillations between the opposite legs (see Figure~\ref{fig:tdisv0.01}(b)). This indicates a transition of the initial wave pulse from propagating into standing mode after several reflections. By examining the phase relationship between velocity and density oscillations for a location on the leg (not shown but similar to Figure~\ref{fig:prf1v0.01}(a)), we observe that the phase shift becomes about a quarter-period after $t=230 \tau_A$. Meanwhile, the density oscillation at the apex shows that the amplitudes do not decrease to nearly zero until $t=260\tau_A$ (see Figure~\ref{fig:prf1v0.01}(c)). This indicates that a standing mode completely forms at $t\approx 260\tau_A$, taking about $2.5 P$ from the launch of the flow pulse. 

\begin{figure}
\centerline{\includegraphics[width=1.0\textwidth,clip=]{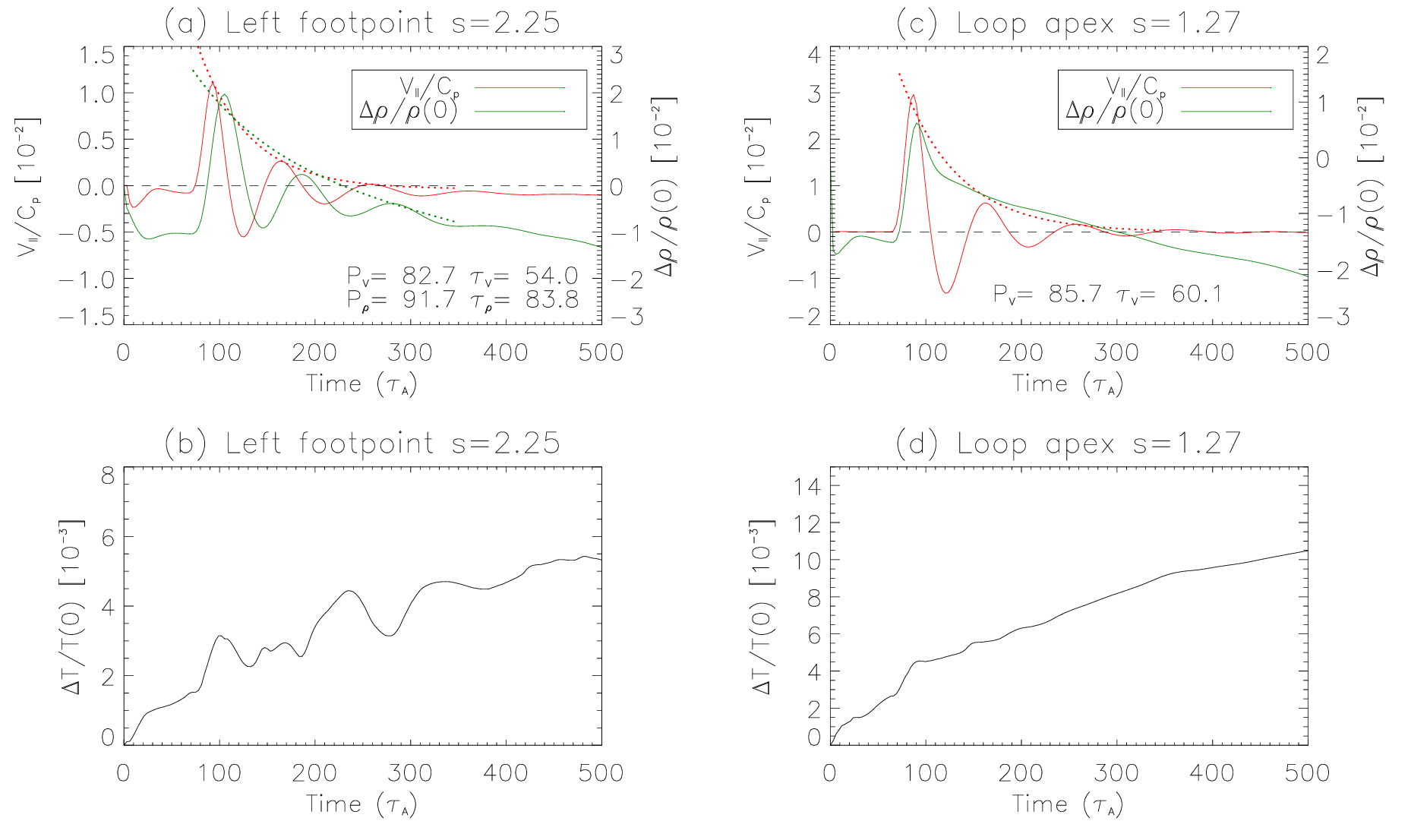}}
\caption{ Simulation results for Case 2B with $A_0=0.01$. The annotations for the lines in each panel are the same as those in Figure~\ref{fig:prf1v0.01}. }
\label{fig:prf2v0.01}
\end{figure}

We then measure the oscillation period and damping time for velocity, density, and temperature perturbations at the remote footpoint and apex of the loop using the time profiles shown in Figure~\ref{fig:prf1v0.01}. By comparing the results for Case 1B with Case 1 (see Table~\ref{tab:ptd}), we find that their differences are small except for density and temperature perturbations at the apex. The oscillation $Q$-factor for velocity, density, and temperature perturbations at the footpoint in Case 1B is only about 10\% larger on average than that in Case 1. The $Q$-factor measured for velocity perturbation at the loop's apex in Case 1B is slightly larger compared to Case 1, similar to the result for the footpoint. However, the $Q$-factors for density and temperature perturbations measured at the apex are much smaller in Case 1B, approximately half of those in Case 1, which is consistent with the gradual formation of a density/temperature node in the loop's apex. 

The differences analyzed in the comparison between Case 1 and Case 1B suggest that nonlinearity has a minor impact on wave damping but can significantly influence the excitation time of a standing wave in coronal loops through impulsive heating when using classical transport coefficients. Specifically, nonlinearity causes a delay in the standing wave formation (see the discussion in Section~\ref{sct:dc}).

In Case 2B, we observe the immediate formation of a standing wave after a single reflection of the initial propagating pulse, in agreement with Case 2. Measurements of wave properties (see Table~\ref{tab:ptd}) indicate that the oscillation $Q$-factors for velocity and density perturbation at the footpoint, and for velocity at apex of the loop, agree well between Case 2B and Case 2. This result suggests that the nonlinearity effect in Case 2 is completely suppressed due to the significant enhancement of compressive viscosity.

Another noticeable feature observed is the prominent increasing trend of the background temperature due to viscous heating in Case 2B (see Figure~\ref{fig:tdisv0.01}(f) and Figures~\ref{fig:prf2v0.01}(c) and (d)). A quantitative comparison between Case 2B and Case 2 indicates that their maximum increases in relative amplitude ($\Delta{T}/T(0)$) are indeed comparable.

\subsection{Effects of Transverse Structuring in the Loop}
\label{sct:ets}
To investigate the influence of loop transverse non-uniformity on the evolution of excited waves in Cases 1 and 2, we conduct three control numerical experiments. In the first control numerical experiment (Case C), we set the loop's peak density ratio ($\chi_\rho=\rho_{\rm in}/\rho_{\rm ex}$) to 1, while maintaining its peak temperature ratio ($\chi_T=T_{\rm in}/T_{\rm ex}$) at 1.5. In the second control numerical experiment (Case D), we set $\chi_T=1$, but retain $\chi_\rho=1.5$. In the third control numerical experiment (Case E), we set $\chi_T=2$, while keep $\chi_\rho=1.5$.

\begin{figure}
\centerline{\includegraphics[width=1.0\textwidth,clip=]{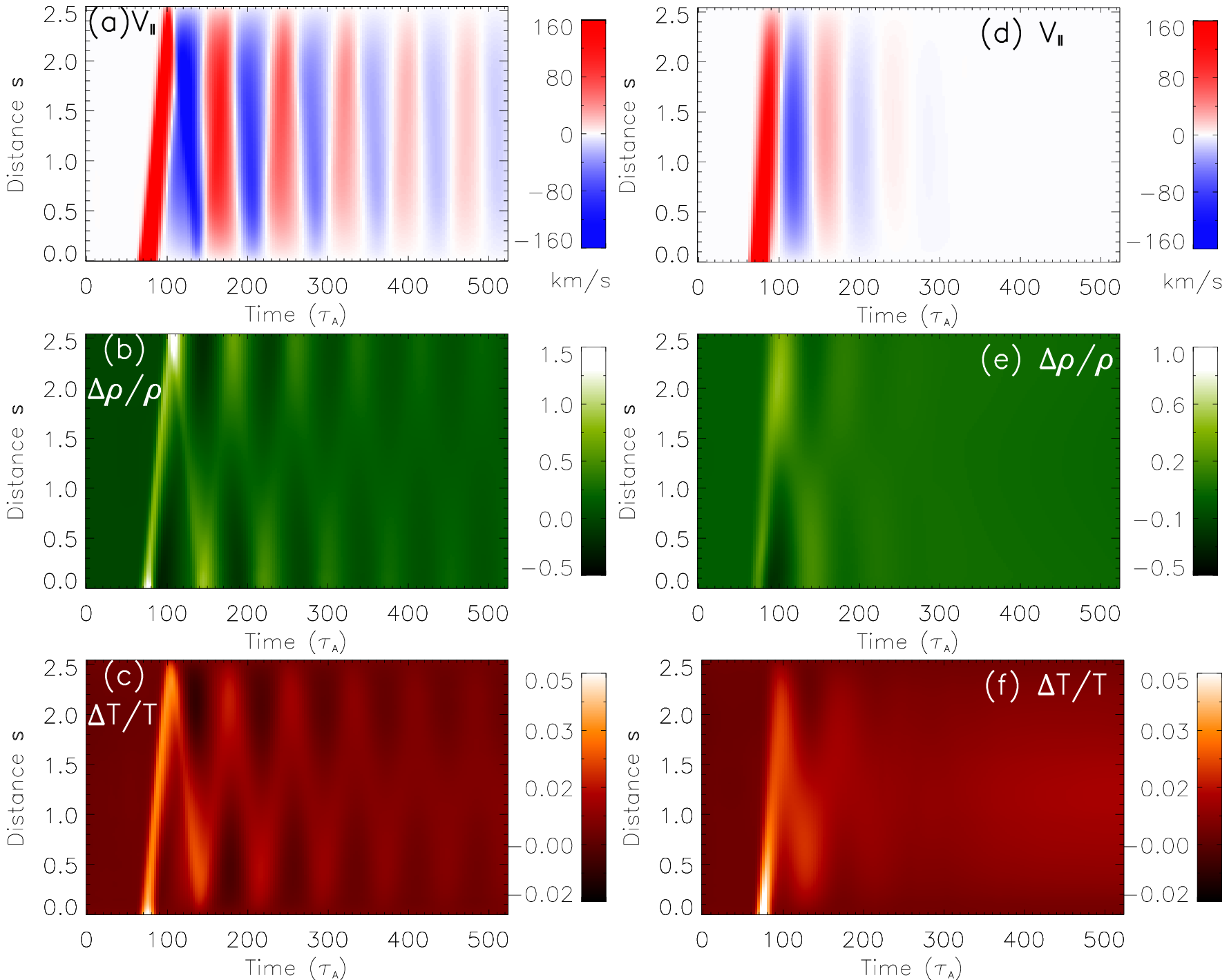}}
\caption{  
Simulation results for Case C with the loop density contrast $\chi_\rho=1.0$ and temperature contrast $\chi_T=1.5$. \textit{Left panels:} Time distance maps of (a) the field-parallel velocity, (b) the relative density perturbation, and (c) the relative temperature perturbation for a slice along the loop for Case 1C. \textit{Right panels:} Same as the left panels but for Case 2C.}
\label{fig:tdisnr1}
\end{figure}

\begin{figure}
\centerline{\includegraphics[width=1.0\textwidth,clip=]{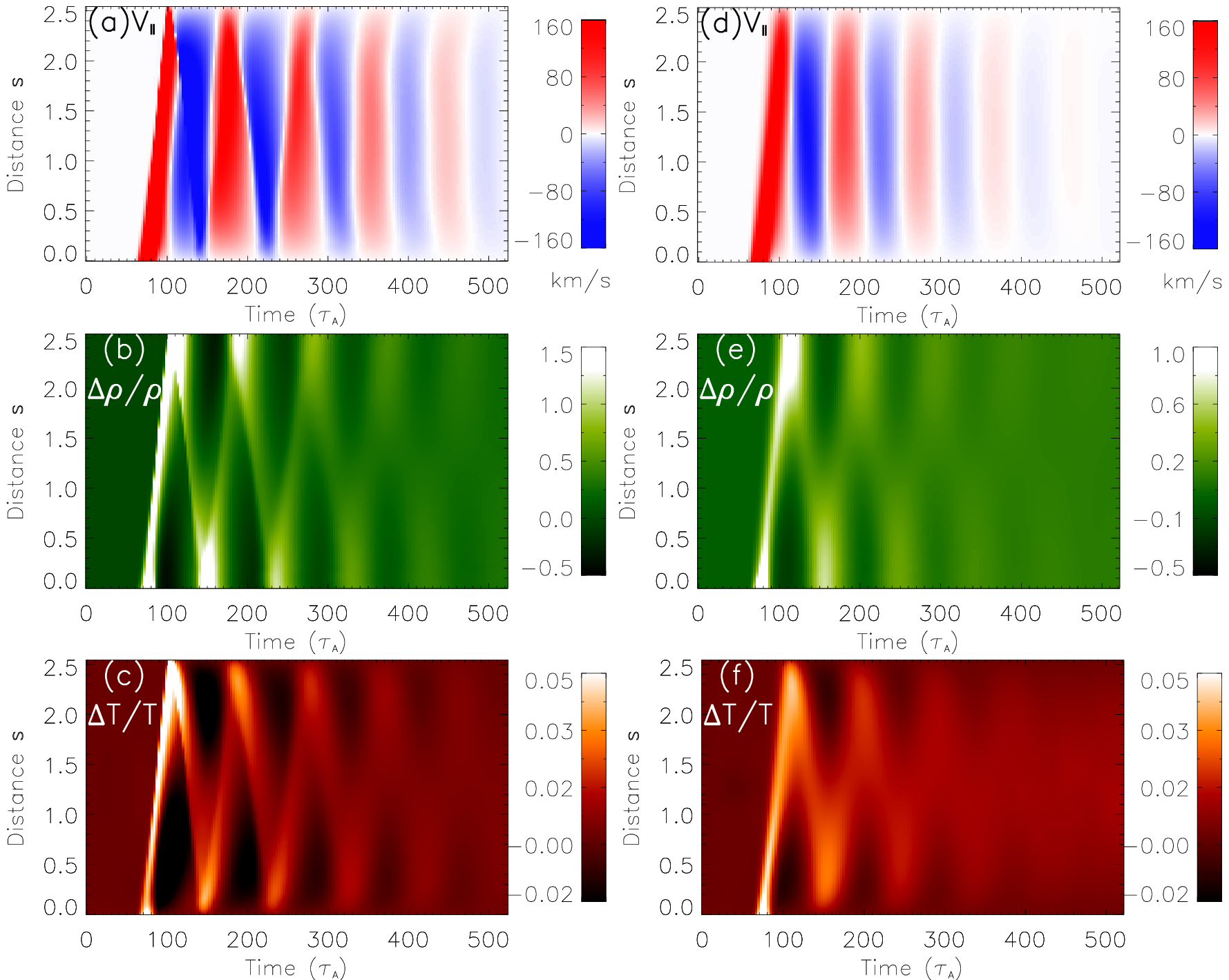}}
\caption{  
Simulation results for Case D with the loop density contrast $\chi_\rho=1.5$ and temperature contrast $\chi_T=1.0$. \textit{Left panels:} Time distance maps of (a) the field-parallel velocity, (b) the relative density perturbation, and (c) the relative temperature perturbation for a slice along the loop for Case 1D. \textit{Right panels:} Same as the left panels but for Case 2D.}
\label{fig:tdistr1}
\end{figure}

\begin{figure}
\centerline{\includegraphics[width=1.0\textwidth,clip=]{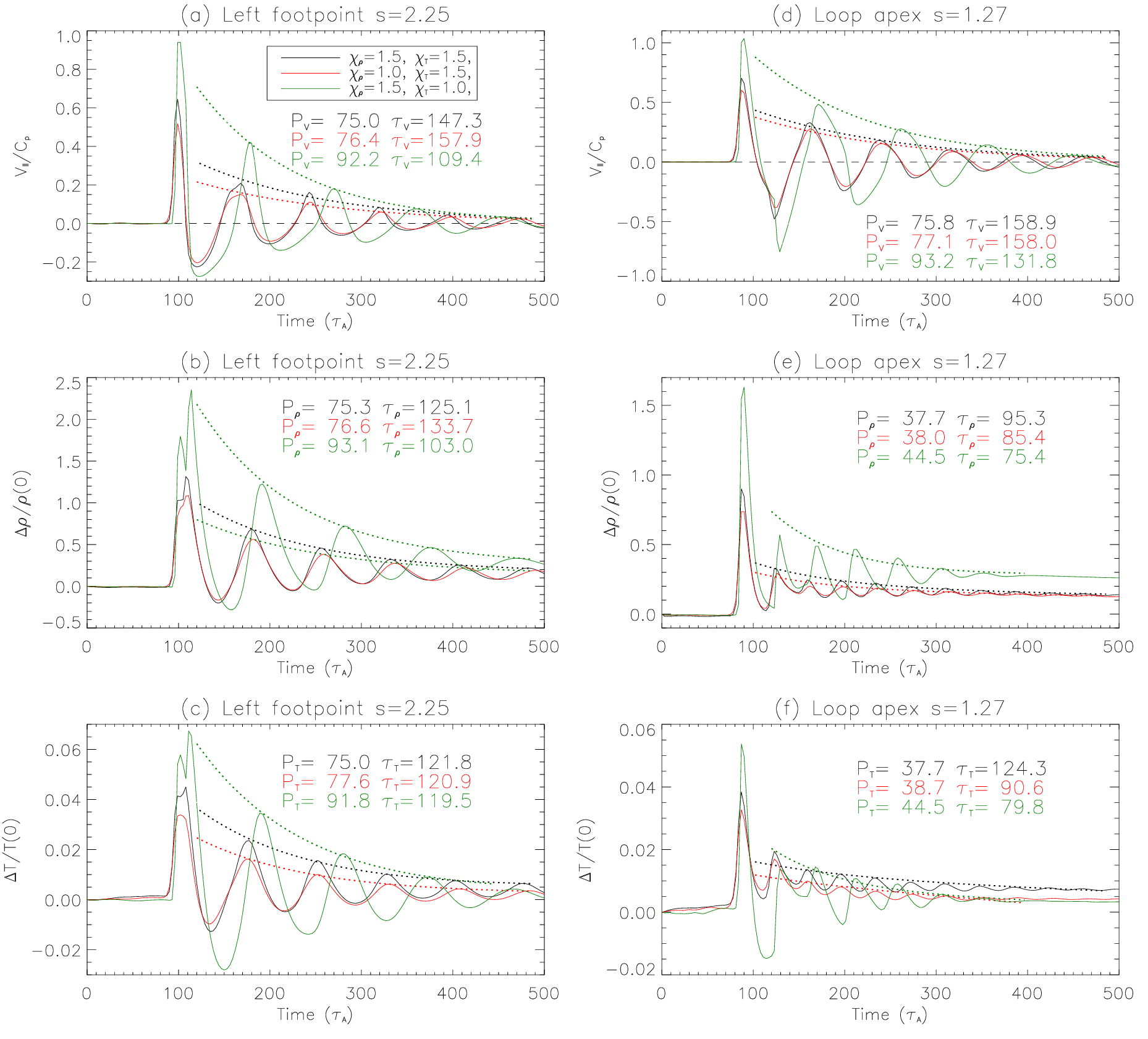}}
\caption{ Comparison of simulations between Case 1 ($\chi_\rho$=1.5 and $\chi_T$=1.5), Case 1C ($chi_\rho$=1.0 and $\chi_T$=1.5), and Case 1D ($\chi_\rho$=1.5 and $\chi_T$=1.0). Temporal evolution of (a) the field-parallel velocity component relative to the local sound speed, (b) the relative density perturbation, and (c) the relative temperature perturbation at the location $s$=2.25 near the left footpoint of the loop. The solid curves in black, red, and green correspond to cases 1, 1C, and 1D, respectively. The dotted lines indicate the exponential decay time fit. (d)-(f): Same as (a)- (c) but for the case at the location $s$=1.27 near the loop's apex. }
\label{fig:prfcmp1}
\end{figure}

\begin{figure}
\centerline{\includegraphics[width=1.0\textwidth,clip=]{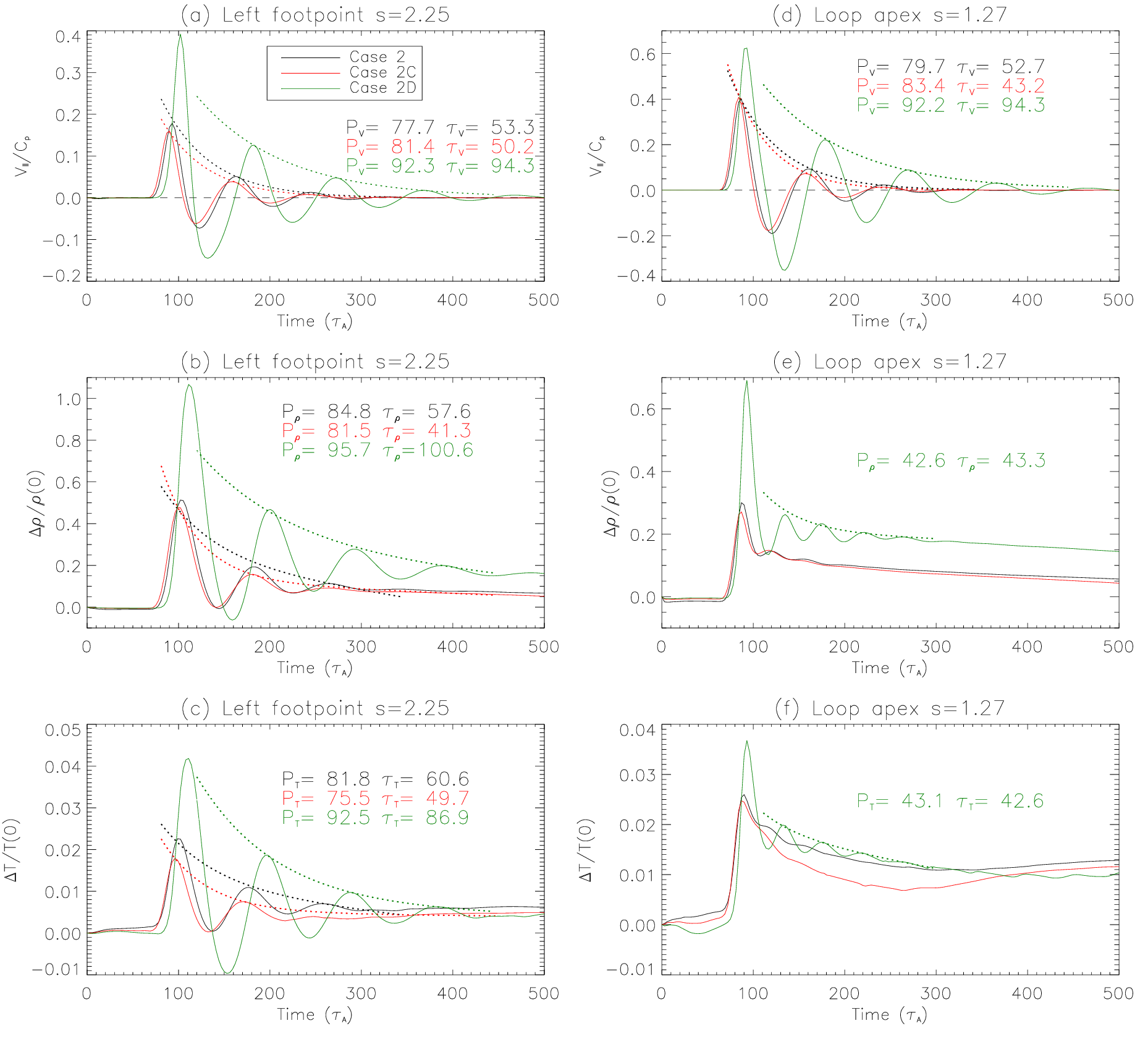}}
\caption{ Comparison of simulations between Case 2 ($\chi_\rho$=1.5 and $\chi_T$=1.5), Case 2C ($\chi_\rho$=1.0 and $\chi_T$=1.5) and Case 2D ($\chi_\rho$=1.5 and $\chi_T$=1.0). Annotations for the lines in each panel are the same as in Figure~\ref{fig:prfcmp1}. }
\label{fig:prfcmp2}
\end{figure}

Figure~\ref{fig:tdisnr1} presents a comparison of time-distance maps for the magnetic field-parallel velocity component (Top Panels), relative perturbed density (Middle Panels), and relative perturbed temperature (Bottom Panels) between Cases 1C and 2C. Similarly, Figure~\ref{fig:tdistr1} compares the time-distance maps between Cases 1D and 2D. In Figure~\ref{fig:prfcmp1}, we compare the time evolution of these physical quantities at the loop's left footpoint and apex for Cases 1, 1C, and 1D. Additionally, in Figure~\ref{fig:prfcmp2}, we perform a similar comparison for Cases 2, 2C, and 2D.   

The time distance maps reveal that there is almost no difference in the wave behavior between Case 1 (2) and Case 1C (2C). From the time profiles at the loop's footpoint and apex, we measure the wave period, damping time, and maximum amplitude for the velocity, density, and temperature perturbations for Cases 1C and 2C. We find that the amplitudes of the excited waves in Cases 1C and 2C are slightly lower (on average by $\sim$26\%) than in Cases 1 and 2. The oscillation $Q$-factor in Case 1C is nearly the same as that in Case 1, while the $Q$-factor in Case 2C is slightly smaller (on average by $\sim$21\%) than that in Case 2. The small decreases in wave amplitude and $Q$-factor could be attributed to the nearly absent transverse density structure in Case C (as discussed in Section~\ref{sct:isp}), which may slightly increase the leakage effect of slow waves in the loop. 

From the comparison of the time distance maps of Cases 1D and 2D with Cases 1 and 2, we observe some distinct differences. Over a simulation time from $t$=60 to 520 $\tau_A$, the wave is reflected back and forth in the loop over five cycles in Case 1D compared to six cycles in Case 1. The damping of the waves excited in Case 2D is noticeably slower than in Case 2. Similarly, the formation speed of standing waves in Case 2D is also slower compared to Case 2. From the time profiles at the loop's footpoint and apex, we measure the wave period, damping time, and maximum amplitude for the velocity, density, and temperature perturbations in Cases 1D and 2D. We find that the amplitudes of the excited waves in Cases 1D are much larger (on average by a factor of 2.2) than in Cases 1, while the amplitudes in Case 2D are larger on average by a factor of 2.9 than in Cases 2.

It is evident from the time profiles, as shown in Figures~\ref{fig:prfcmp1} and \ref{fig:prfcmp2}, that the wave periods ($P_D$) for Cases 1D and 2D are longer than those ($P$) for Cases 1 and 2. We estimate their mean ratio, $\overline{P_D/P}=1.19\pm0.05$, from the measurements of wave periods listed in Table~\ref{tab:ptd}. This difference can be explained by the dependence of wave propagation speed on the plasma temperature in the loop, $C_p\propto T^{1/2}$ \citep[see][]{wan21}. Consequently, from the equation for wave period, $P=2L/C_p$, we obtain $(P_D/P)_{\rm the}\propto (T_m/T_{mD})^{1/2}\approx1.22$, where $T_m$ and $T_{mD}$ are the maximum temperature in the loop for Cases 1 and 2 and for Cases 1D and 2D, respectively, and $T_m/ T_{mD}=1.5$. Thus, the prediction by linear theory agrees well with the simulated result. 

From the measurements listed in Table~\ref{tab:ptd}, we find that the average $Q$-factor for Case 1D is approximately $1.4\pm0.3$, while for Case 1 it is about $2.2\pm0.6$. This indicates that the wave damping in Case 1D is faster than in Case 1, with a ratio $\overline{Q_{1D}/Q_1}\approx0.64$. On the other hand, for Case 2D, the average $Q$-factor is around $1.00\pm0.06$, whereas for Case 2, it is about $0.73\pm0.05$. This suggests that the wave damping in Case 2D is slower than in Case 2, with a ratio $\overline{Q_{2D}/Q_2}\approx1.37$. 

These distinct differences in damping characteristics between Cases 1D and 2D, when compared to Cases 1 and 2, suggest that the waves excited in Model 1 and Model 2 may be dominated by different damping mechanisms. As discussed in Section~\ref{sct:etc}, we have determined that thermal conduction damping is negligible due to the assumption of $\gamma=1.05$ in this study. Therefore, we consider compressive viscosity and wave leakage as the two major damping mechanisms. The oscillation $Q$-factor for slow waves, including dissipation by compressive viscosity alone, can be derived from the dispersion relation \citep[refer to][]{wan21}:
\begin{equation}
     Q=\frac{\tau}{P} = \frac{3}{8\pi^2}\left(\frac{1}{\epsilon}\right) \propto\frac{n_0L}{T_0^2},
  \label{eq:qvs}   
\end{equation}
where the viscous ratio, $\epsilon$, is defined as
\begin{equation}
 \epsilon=\frac{1}{R}=\frac{\eta_0}{\rho_0 C_s^2 P_0},
 \label{equ:vis}
\end{equation}
where $R$ is the Reynolds number, and $P_0=2L/C_s$. Thus, the ratio of $Q$-factors between the two cases with a temperature ratio of $T_m/T_{mD}$=1.5 can be estimated as $(Q_D/Q)_{\rm the}=(T_m/T_{mD})^2$=2.25. The measured result of $\overline{Q_{2D}/Q_2}>1$ for Case 2D supports this prediction, suggesting that in the case when the compressive viscosity coefficient is significantly enhanced, the damping of simulated waves is dominated by viscous dissipation. The fact that $\overline{Q_{2D}/Q_2}<(Q_D/Q)_{\rm the}$ can be explained by the additional damping due to wave leakage. Whereas the result of  $\overline{Q_{1D}/Q_1}<1$ for Case 1D, inconsistent with the prediction, suggests that wave leakage may dominate over compressive viscosity in wave damping in the case with no temperature enhancement within the loop and in a normal coronal condition with classical viscosity coefficient. On the other hand, this result implies that with the increase of temperature contrast from $\chi_T=1$ to 1.5, the leakage effect should reduce (or the hotter structure facilitates the trapping of excited waves in the loop). 

To further investigate the relative influence of coronal leakage and compressive viscosity on wave damping in a hot coronal loop, we conducted simulations in Case E with a temperature contrast $\chi_T=2$ (not shown). We measured the wave period and damping time using the same methodology as in the other cases and compiled the measured parameters in Table~\ref{tab:ptd}. 

We observed that for Case 1E, the average oscillation $Q$-factor is around 1.9, which is only slightly smaller than that ($Q_1\approx2.2$) of Case 1. Notably, the average $Q$-factor is computed from values of $Q$-factor for velocity, density, and temperature perturbations at the footpoint, as well as for velocity at the apex. This computation excludes values of $Q$-factor for density and temperature perturbations at the apex due to their larger measurement uncertainties. Drawing from the analogous discussion as presented earlier, we would anticipate that $(Q_E/Q)_{\rm the}=0.56$ for $T_m/T_{mE}=1.5/2.0$ if viscous damping were the predominant factor. However, our simulation results reveal that $\overline{Q_{1E}/Q_1}\approx0.9$. This suggests that the increase in viscous damping with temperature is offset by an enhanced trapping effect (or weakened wave leakage). This, in turn, indirectly indicates that the damping rate induced by wave leakage is of a similar magnitude to that caused by compressive viscosity. Consequently, this observation highlights the potential significance of wave leakage in comprehending the mechanisms governing wave damping in hot coronal loops when conventional transport coefficients are considered.

In Case 2E, we observed that the average $Q$-factor is approximately 0.45, indicating a decrease compared to the value of $Q_2=0.73$ for Case 2. Notably, the observed ratio, $\overline{Q_{2E}/Q_2}=0.62$, closely aligns with the predicted ratio $(Q_E/Q)_{\rm the}=0.56$. This alignment further reinforces the idea that compressive viscosity plays a significant role as the dominant damping mechanism in Model 2, a crucial factor in comprehending the rapid excitation of standing slow-mode waves observed in hot coronal loops.

\begin{sidewaystable}
	\caption{Measured parameters for simulated loop oscillations in $B$-parallel velocity, density, and temperature for different cases. Items marked with `--' indicates cases with no measurable oscillations. The oscillation periods ($P$) and damping times ($\tau$) are in units of the Alfv\'{e}n time ($\tau_A$).}
\label{tab:ptd}
\begin{tabularx}{\textheight}{lccccccccccccccccccc} 
 \hline
   & \multicolumn{9}{c}{Left footpoint} & & \multicolumn{9}{c}{Loop apex} \\
    \cline{2-10} \cline{12-20} \\
  Case & $P_V$ & $\tau_V$ & $\left(\frac{\tau}{P}\right)_V$ & $P_\rho$ & $\tau_\rho$ & $\left(\frac{\tau}{P}\right)_\rho$ & $P_T$ & $\tau_T$ & $\left(\frac{\tau}{P}\right)_T$ & & $P_V$ & $\tau_V$ & $\left(\frac{\tau}{P}\right)_V$ & $P_\rho$ & $\tau_\rho$ & $\left(\frac{\tau}{P}\right)_\rho$ & $P_T$ & $\tau_T$ & $\left(\frac{\tau}{P}\right)_T$\\
 \hline
%  0 & 77 & 181 & 2.4 & 75 & 154 & 2.1 & 75 & 161 & 2.2 && 74 & 187 & 2.5 & 37 & 174 & 4.7 & 37 & 206 & 5.5 \\
% \hline
 1  & 75 & 147 & 2.0 & 75 & 125 & 1.7 & 75 & 122 & 1.6 && 76 & 159 & 2.1 & 38 & 95 & 2.5 & 38 & 124 & 3.3 \\
 1A & 75 & 147 & 2.0 & 75 & 125 & 1.7 & 74 & 131 & 1.8 && 76 & 159 & 2.1 & 38 & 95 & 2.5 & 38 & 156 & 4.1 \\
 1B & 79 & 163 & 2.1 & 75 & 162 & 2.2 & 75 & 126 & 1.7 && 77 & 174 & 2.3 & 42 & 51 & 1.2 & 42 & 63 & 1.5 \\
 1C & 76 & 158 & 2.1 & 77 & 134 & 1.8 & 78 & 121 & 1.6 && 77 & 158 & 2.1 & 38 & 85 & 2.3 & 39 & 91 & 2.3 \\
 1D & 92 & 109 & 1.2 & 93 & 103 & 1.1 & 92 & 120 & 1.3 && 93 & 132 & 1.4 & 45 & 75 & 1.7 & 45 & 80 & 1.8 \\
 1E & 71 & 140 & 2.0 & 71 & 118 & 1.7 & 71 & 125 & 1.8 && 71 & 143 & 2.0 & 35 & 70 & 2.0 & 31 & 20 & 0.6 \\
 \hline
 2  & 78 & 53 & 0.7 & 85 & 58 & 0.7 & 82 & 61 & 0.8 && 80 & 53 & 0.7 & \multicolumn{3}{c}{------------} &  \multicolumn{3}{c}{------------}\\
 2A & 78 & 53 & 0.7 & 85 & 58 & 0.7 & 74 & 79 & 1.1 && 80 & 53 & 0.7 & \multicolumn{3}{c}{------------} &  \multicolumn{3}{c}{------------}\\
 2B & 83 & 54 & 0.7 & 92 & 84 & 0.9 & \multicolumn{3}{c}{------------} && 86 & 60 & 0.7 & \multicolumn{3}{c}{------------} & \multicolumn{3}{c}{------------}\\
 2C & 81 & 50 & 0.6 & 82 & 41 & 0.5 & 76 & 50 & 0.7 && 83 & 43 & 0.5 & \multicolumn{3}{c}{------------} & \multicolumn{3}{c}{------------}\\
 2D & 92 &  94 & 1.0 & 96 & 101 & 1.1 & 93 & 87 & 0.9 && 92 & 94 & 1.0 & 43 & 43 & 1.0 & 43 & 43 & 1.0\\
 2E  & 71 & 38 & 0.5 & 79 & 23 & 0.3 & 73 & 35 & 0.5 && 74 & 37 & 0.5 & \multicolumn{3}{c}{------------} &  \multicolumn{3}{c}{------------}\\
 \hline
\end{tabularx}
\end{sidewaystable}

\subsection{Effects of Wave Leakage and Mode Coupling}
\label{sct:emc}

\begin{figure}
\centerline{\includegraphics[width=1.0\textwidth,clip=]{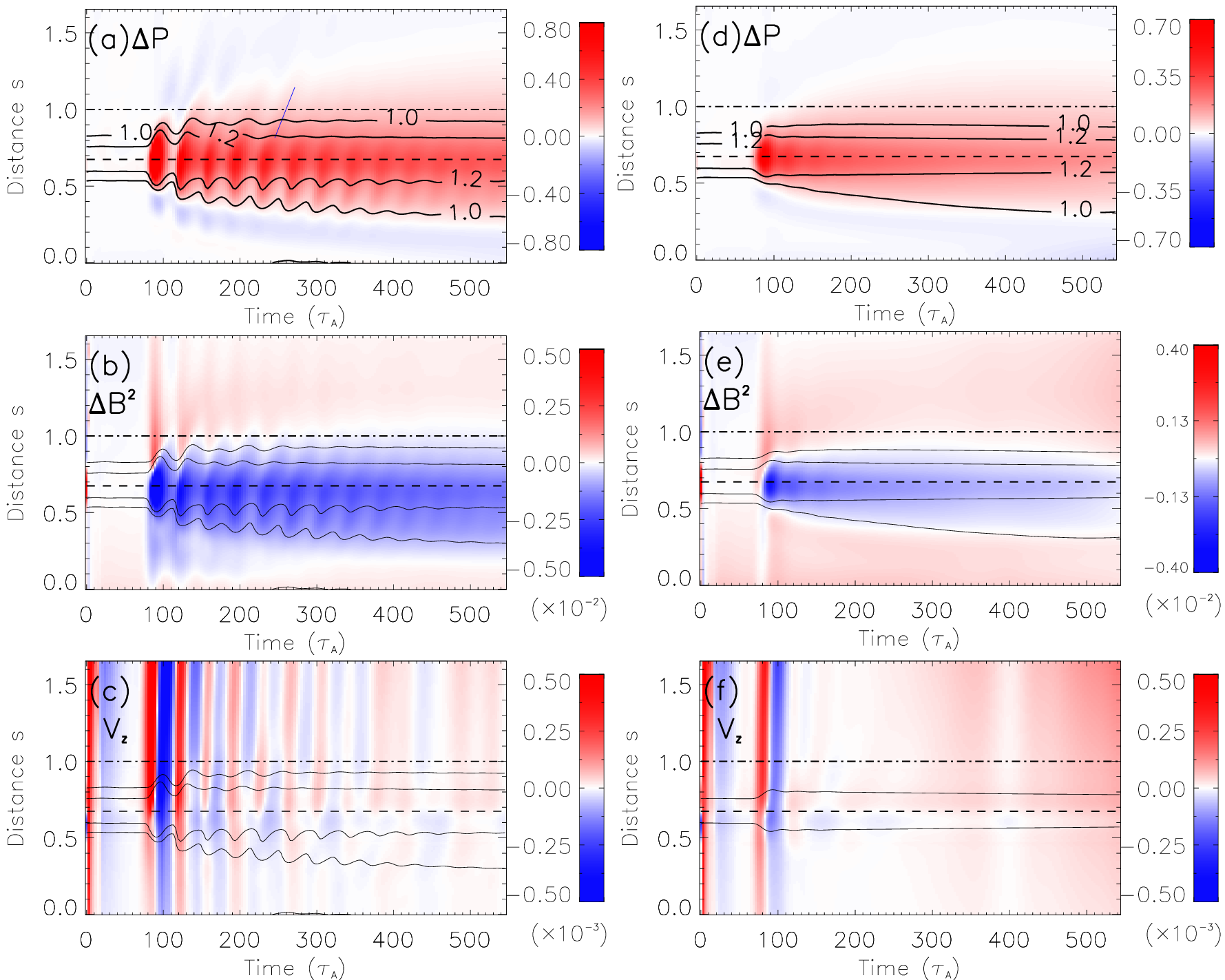}}
\caption{{\it Left panels:} Time-distance maps representing (a) perturbed gas pressure ($\Delta{P}=P(t)-P(6\tau_A)$), (b) perturbed magnetic pressure ($\Delta{B^2}=B^2(t)-B^2(6\tau_A)$), and (c) the $z$-component of velocity ($V_z$) along a vertical cut at the loop's apex in Case 1. {\it Right panels:} Similar to the left panels, but for Case 2. The distance $s$ is measured from the lower end of the cut. In each panel, solid contours delineate the loop boundaries with $\rho=$ 1.0 and 1.2, while dashed and dot-dashed lines indicate the locations for time profile plots shown in Figure~\ref{fig:vtpf}. A blue solid line in (a) indicates the measurement of upwardly propagating perturbations with a speed of 0.013 $V_A$.  }
\label{fig:vtds}
\end{figure}

We analyze the nonlinear effects related to wave leakage and mode coupling by studying the evolution of simulated waves across the loop. In Figure~\ref{fig:vtds}, we present time-distance maps illustrating gas pressure perturbations ($\Delta{P}=\Delta({\rho}T)$), magnetic pressure perturbations ($\Delta{B^2})$, and the $z$-component of velocity along a cut across the loop apex (see Figure~\ref{fig:init}(a)). The perturbations, $\Delta{P}(t)=P(t)-P(t_0)$ and $\Delta{B^2}(t)=B^2(t)-B^2(t_0)$ at a given time $t$, are calculated in relation to a reference time $t_0=6\tau_A$, as the initial state requires some relaxation time to reach equilibrium. By defining the loop boundaries with $\rho$=1.0 and 1.2, we observe that gas and magnetic pressure perturbations inside the loop are in anti-phase, consistent with the characteristics of slow-mode waves. In Case 1, which uses classical transport coefficients, the enhancement in gas pressure, as it passes the loop apex, leads to significant variations in the loop boundary. The expansion of the loop triggers the excitation of fast-mode waves propagating upward. This is evidenced by enhancements in magnetic pressure outside the loop, which are in phase with the velocity component $V_z$ (see Panels (b) and (c)). This provides clear evidence for mode coupling due to nonlinearity. We also notice that during the time $90<t<200$, the plasma within the loop displays the coherent up and down motions from the evolution of $V_z$ as shown in Panel (c), suggesting the coupling of slow waves with kink oscillations of the loop. 

Note that the induced vertical kink oscillation exhibits a period identical to that of the density oscillation at the loop apex, or half the period of the slow-mode waves. In contrast, the period of the fundamental kink mode is estimated to be about $10\tau_A$ (approximately $P/8$ of the slow waves). This indicates that the observed kink oscillation is driven by the slow waves rather than the free mode excited by an initial flow pulse, as predicted by \citet{koh18}. It is important to mention that there is energy transfer from the slow to kink mode due to the (slight) transverse displacement of the loop by the centrifugal force. This is a nonlinear process, dependent on the square of the velocity amplitude, $V_\|^2$. A similar mode-coupling effect was also observed in the 3D MHD modeling of slow-mode waves excited by an impulsive onset of steady inflow by \citet{ofm12} (see their Figures 5 and 8). 

Furthermore, we have observed that the gas pressure enhancements, except for the initial pulse, are linked to weak and slowly ascending features outside the loop. These features are estimated to propagate at approximately 0.013 $V_{A0}$ or 0.26 $C_{s0}$ (see Figure~\ref{fig:vtds}(a)). Upon closer examination through animations, we have determined that these slowly propagating features result from wavefront extensions outside the loop, which delay in phase with heights. These leaked waves may stem from strong nonlinear interactions of flows at the footpoints.  Notably, the leaked wavefront is not associated with the initial pulse during its propagation. The phase delay of this leaked wavefront can be explained by the fact that these waves are nearly simultaneously generated during the reflection at the footpoint. They propagate along field lines of varying sizes, with the wave along the outer field line taking more time to reach the vertical cut compared to the inner one, as their propagation speeds are close.    

In Case 2, where we employ viscosity enhanced by a factor of 10, we observe wave behaviors that are notably distinct from those in Case 1. Specifically, we notice that only the initial gas pressure pulse causes a noticeable expansion of the loop and excites outward propagating fast-mode waves and the loop kink motions through mode coupling (Figures~\ref{fig:vtds}(d)-(f)). Additionally, there are nearly no indications of leaked waves propagating outside the loop, associated with gas pressure enhancements within the loop. These disparities indicate that nonlinear effects in Case 2 are effectively suppressed, mainly due to the significantly enhanced viscosity. Consequently, wave leakage in this scenario is substantially reduced in comparison to Case 1. This finding provides direct support to the conclusion drawn in the previous section.

\begin{figure}
\centerline{\includegraphics[width=1.0\textwidth,clip=]{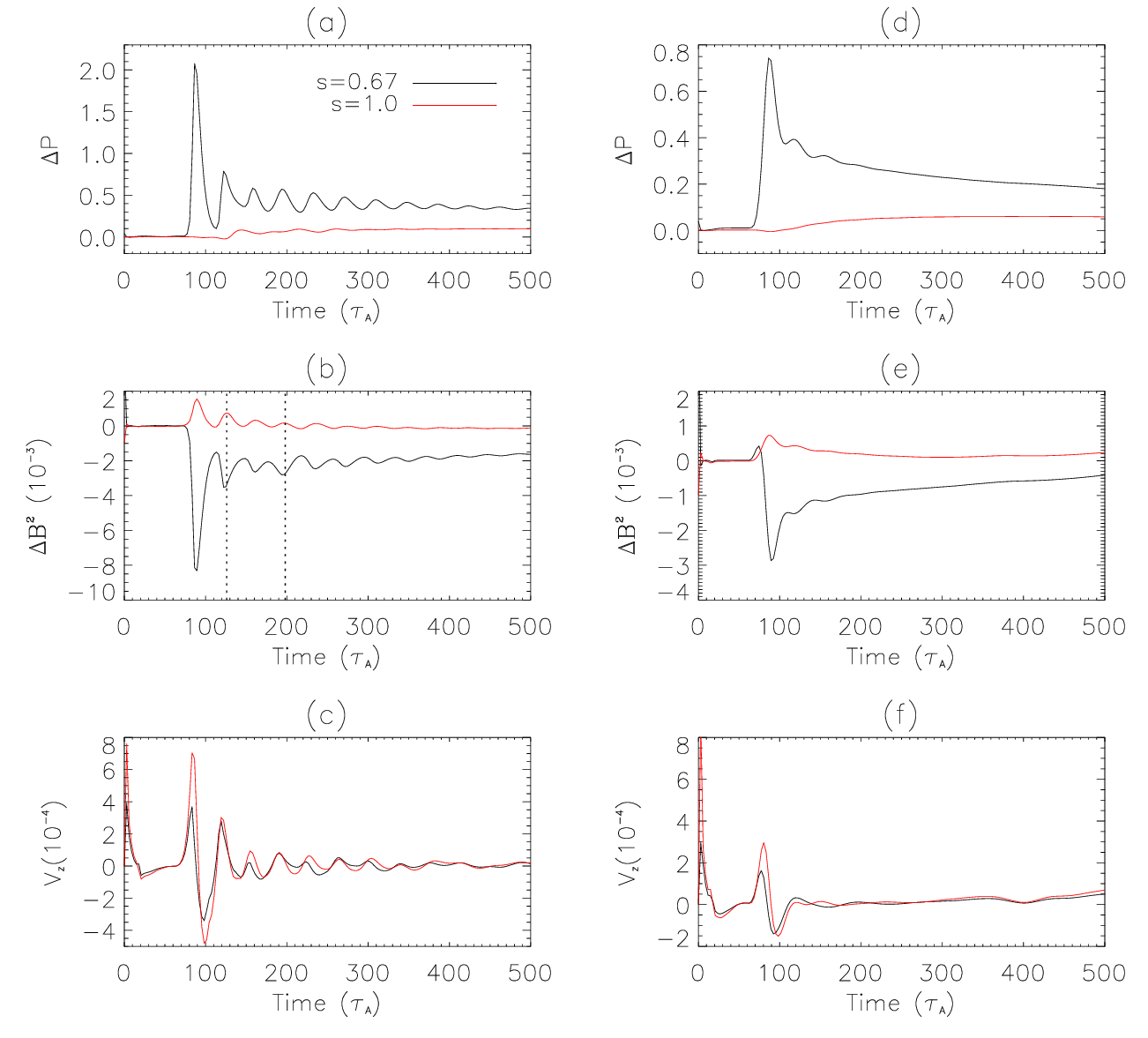}}
\caption{ Temporal evolution of (a) perturbed gas pressure ($\Delta{P}$), (b) perturbed magnetic pressure ($\Delta{B^2}$), and  (c) the $z$-component of velocity ($V_z$) at two locations along the vertical cut for Case 1. The black curves stand for the location at $s=0.67$ inside the loop, while the red curves for $s=1.0$ above the loop's apex. The vertical dotted lines in (b) illustrate the time delay of perturbations $\Delta{B^2}$ between $s=0.67$ and $s=1.0$. Panels (d)-(f) depict the same quantities as (a)-(c) but for Case 2.}
\label{fig:vtpf}
\end{figure}

In Figure~\ref{fig:vtpf}, we compare the temporal variations of gas pressure perturbation, magnetic pressure perturbation, and the $z$-component of velocity at two positions along the vertical cut. One position is located at the center of the loop with $s$=0.67, and the other is situated above the loop with $s$=1.0, as indicated in Figure~\ref{fig:vtds}. In Case 1, we measure the average amplitudes of gas and magnetic pressures inside the loop during the time $75<t<500$ to be $\overline{\Delta{P}}=0.43$ and $\overline{\Delta{B^2}}=-2.2\times10^{-3}$, respectively. The ratio of changes in the dimensional gas and magnetic pressures can be estimated as $\Delta{P_g}/\Delta{P_m}=\beta_0(\overline{\Delta{P}}/\overline{\Delta{B^2}})=-0.96$, where the plasma $\beta_0=0.0049$, calculated from the normalization parameters (see Table~\ref{tab:para}). This implies that $\Delta{P_g}+\Delta{P_m}\sim{0}$ or $P_g + P_m \sim{\rm const}$. This result corroborates the findings in \citet{ofm22} and aligns with the fundamental property of slow waves, where their total pressure is approximately balanced. Similarly, in Case 2, we estimate $\overline{\Delta{P}}=0.30$ and $\overline{\Delta{B^2}}=-1.0\times10^{-3}$ from the averages during  $75<t<350$. This leads to $\Delta{P_g}/\Delta{P_m}=-1.4$. Here, we observe that the amplitude of gas pressure perturbations is slightly larger than that of magnetic pressure perturbations. Given that gas pressure is the primary restoring force in slow waves, this suggests that the coupling between gas and magnetic pressures in Case 2 is weaker in comparison to Case 1.

Comparisons of time profiles of $V_z(t)$ at the two positions, as depicted in Figures~\ref{fig:vtpf}(c) and (f), reveal a small time shift of approximately $1\tau_A$ between their oscillations. Notably, this time delay also becomes evident when examining the time profiles of $\Delta{B^2}$, particularly when comparing the positive peaks at $s$=1.0 and negative peaks at $s$=0.67 in Figure~\ref{fig:vtpf}(b). Utilizing their separation distance and time delay, we can estimate the upward propagation speed of perturbations in $V_z$ and $\Delta{B^2}$ to be approximately $0.3V_{A0}$, which closely aligns with the initial Alfv\'{e}n speed distribution above the loop top (see Figure~\ref{fig:init}(d)). Given that the fast-mode wave speed in perpendicular propagation $V_f=V_A(1+\gamma\beta/2)^{1/2}\approx{V_A}$ in the low-$\beta$ coronal condition, this observation strongly suggests that these rapidly upward propagating perturbations correspond to fast magnetoacoustic waves.

\subsubsection{Comparison with Observations}
We have selected three longitudinal loop oscillation events that were observed in high-temperature EUV channels using SDO/AIA. These channels include the 94 \AA\ channel, which is sensitive to a temperature of 7 MK, and the 131 \AA\ channel corresponding to 11 MK. While these three events have been previously extensively investigated to understand their trigger and damping mechanisms, there has not been a detailed analysis of how these waves behave across the loop structure. Our goal is to use these observed events to validate the predictions made by our modeling approach regarding the influence of mode coupling and wave leakage, as discussed in the previous section.

These three events include:

(1) An event on 28 December 2013 in AR NOAA 11936, which was associated with a C3.0 flare peaking at 12:47 UT. This event had an oscillation period of 12 minutes and a decay time of 9 minutes, as measured in the 131\AA\ channel \citep{wan15,wan18}. Notably, \citet{wan15} discovered that a standing slow-mode wave emerged after a single reflection of the initial pulse.

(2) An event on 7 May 2012 in AR NOAA 11476, which was associated with a C7.4 flare peaking at 17:26 UT. This event had an oscillation period of 11 minutes and a decay time of 7 minutes in 131 \AA \citep{kum13}. Upon reevaluation, \citet{kri21} unveiled that the initial sloshing oscillations swiftly transitioned into a standing wave after several reflections as observed in the 94 \AA\ channel.  

(3) An event on 20 July 2013 in AR NOAA 11793, which was associated with a C2.1 flare peaking at 3:38 UT. This event exhibited an oscillation period of 7 minutes and a decay time of 19 minutes, measured in the 94 \AA\ channel \citep{kum15}.    

\begin{figure}
\centerline{\includegraphics[width=1.0\textwidth,clip=]{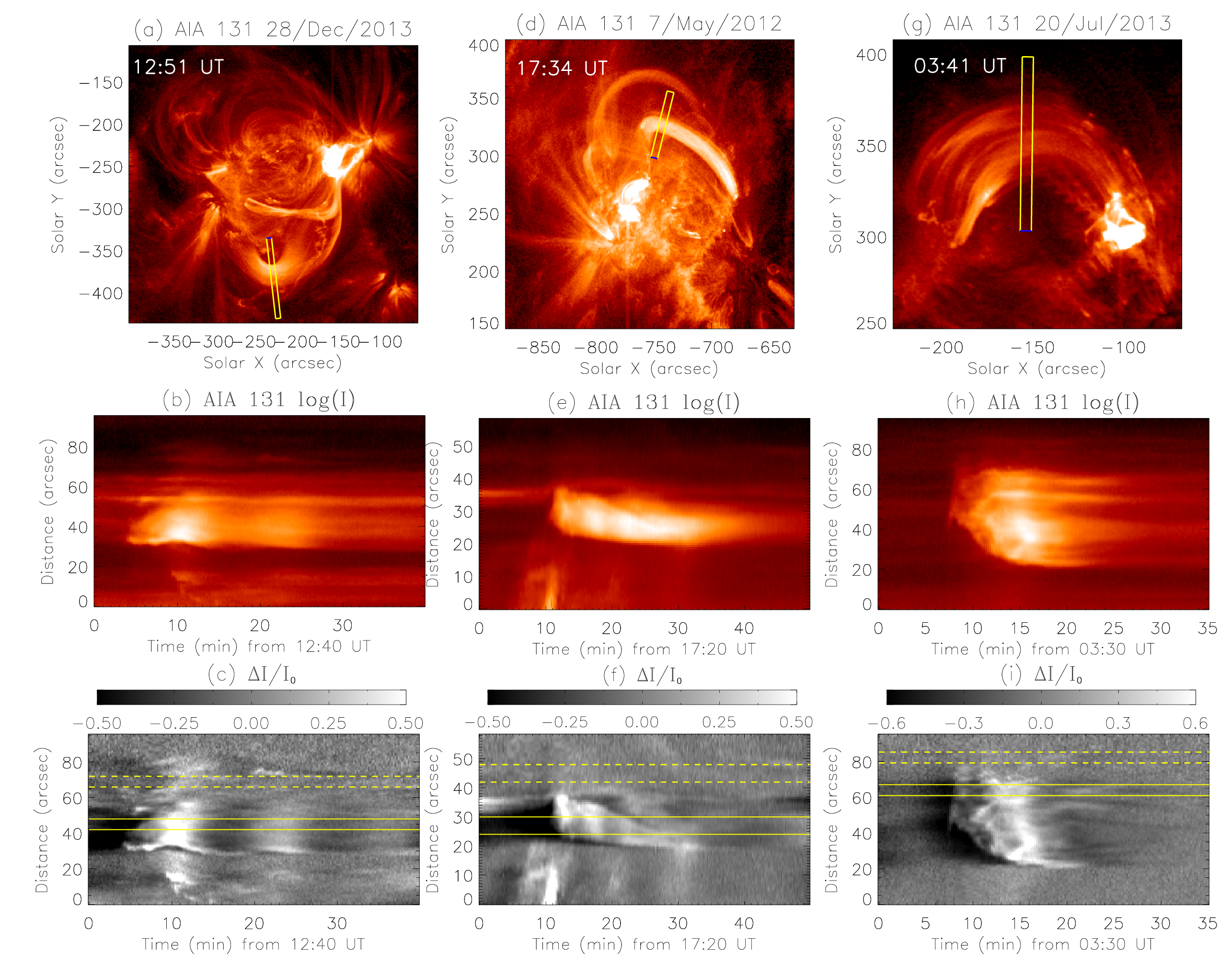}}
\caption{ 
Analysis of wave leakage and mode coupling effects associated with longitudinal loop oscillations observed by SDO/AIA in the 131 \AA\ channel. (a) A snapshot of the event  on 28 December 2013, displaying longitudinal intensity oscillations in a flaring hot loop.  (b) Time-distance map of the 131 \AA\ intensity along a narrow slice across the loop as shown in (a). The distance is measured from the slice's blue end. (c) Time-distance map of the detrended and normalized intensity along the slice. The solid line band indicates a region within the loop, whereas the dashed line band indicates a nearby region. These two regions (11-pixels wide) are used to calculate the time profiles of the detrended intensities shown in Figure~\ref{fig:vlc}.  (d)-(f): The same as (a)-(c) but for the event on 7 May 2012. (g)-(i): The same as (a)-(c) but for the event on 20 July 2013. }
\label{fig:vobs}
\end{figure}

We present snapshots of the analyzed loop structures, captured using the 131 \AA\ channel, in the upper row of Figure~\ref{fig:vobs}. In each event, the observations show that a small flare near one of the loop's footpoints injected hot plasma into the loop, triggering an intensity disturbance that oscillated between the two footpoints. To generate time-distance maps, we sample data  along a slender slice (11 pixels wide) running across the loop's apex and average the emission over its narrow width for a sequence of images. The obtained time-distance maps of 131 \AA\ intensities are shown in the middle row of Figure~\ref{fig:vobs}. To improve the visibility of disturbances, the intensities at each spatial position are detrended and normalized using the formula $(I(t)-I_0(t))/I_0(t)$, where $I_0(t)$ represents a slowly varying trend obtained through Fourier low-pass filtering with a cutoff frequency of approximately twice the wave period. The detrended time-distance maps are displayed in the bottom row of Figure~\ref{fig:vobs}. 

\begin{figure}
\centerline{\includegraphics[width=1.0\textwidth,clip=]{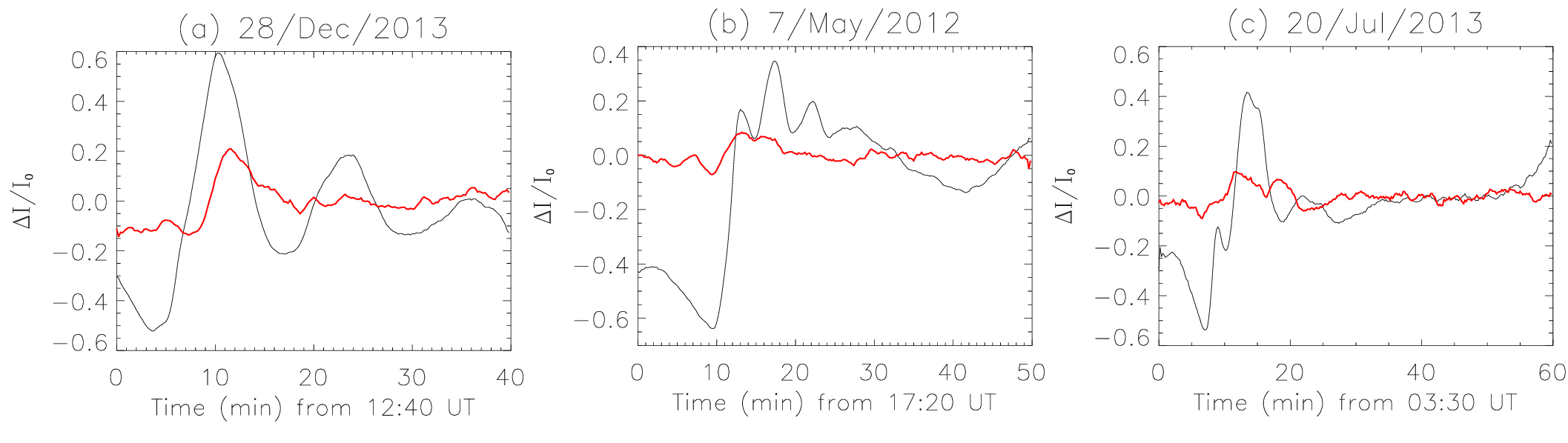}}
\caption{ 
(a) Time profiles of the detrended average intensities in the two regions, as indicated by the bands in Figure~\ref{fig:vobs}(c) for the 28-Dec-2013 event. A 5-pixel smoothing average has been applied to reduce noise. The intensity curve for the area within the loop is in the black line, whereas that for the area near the loop is in the red line.  (b) Similar to (a) but for the two regions shown in Figure~\ref{fig:vobs}(f) for the 7-May-2012 event. (c) Similar to (a) but for the two regions in Figure~\ref{fig:vobs}(i) for the 20-Jul-2013 event. 
}
\label{fig:vlc}
\end{figure}

In all three events, there is no evidence of propagating fast magnetoacoustic waves, which typically propagate at speeds around 1000 km~s$^{-1}$, nor any signs of wave leakage, such as the sub-sonic propagating disturbances that our simulations predicted. However, in the first event, slow waves seem to be associated with the loop's kink oscillations (see Figure~\ref{fig:vobs}(c)). For a more quantitative analysis, we compare the evolution of the detrended average intensities inside and near the loop, as shown in Figure~\ref{fig:vlc}. We observe clear wave signals within the loop, while the external region exhibits only slight enhancements during the initial peak of the waves. These observational features appear to be consistent with the predictions of Case 2, which includes significantly enhanced viscosity. 

It is important to note that in the low-$\beta$ corona, the mode coupling between slow waves and fast-mode waves  is generally weak \citep{dem04b, afa15}. As a result, the excited fast-mode wave due to nonlinear coupling may be too faint to detect by SDO/AIA, as discussed below. According to linear theory for the perpendicular propagation of fast waves, we have the relationship $\rho_1/\rho_0=v_{z1}/v_f\sim v_{z1}/v_A$, where $\rho_1$ represents the dimensional density perturbation, $v_{z1}$ is the dimensional velocity component perpendicular to the magnetic field, and $v_f$ is the fast-mode speed that approximates the local Alfv\'{e}n speed, $v_A$. From the temporal evolution of $V_z$ for Case 1 (see Figure~\ref{fig:vtpf}(c)), we estimate its maximum amplitude, $v_{zm}=7\times 10^{-4}V_{A0}=5$ km~s$^{-1}$. As shown in Figure~\ref{fig:init}(d), the Alfv\'{e}n speed near the loop top is approximately $v_A=0.4V_{A0}$. Based on the assumption that intensity and density perturbations roughly satisfy the relationship $I_1/I_0\sim 2\rho_1/\rho_0$, we can calculate the relative amplitude of intensity perturbations $I_1/I_0\approx0.35\%$. This predicted weak intensity wave signal falls below the detectable threshold of SDO/AIA. However, the corresponding velocity wave signals with an amplitude of $v_{zm}$=5 km~s$^{-1}$ might be detectable through Doppler shifts using imaging spectrometers with very high cadence.    

\section{Discussion and Conclusions}
\label{sct:dc}

Utilizing a nonlinear dissipative 2.5D MHD arcade AR model, which incorporates a hot and dense coronal loop in a gravitationally stratified atmosphere based on observed physical parameters, we have conducted an investigation into the impact of compressive viscosity on the excitation and evolution of reflected propagating and standing slow-mode waves generated by impulsive heating in solar flares. We explore wave properties (e.g., excitation time of the standing mode, wave trapping/leakage, and dissipation) within two distinct loop models: one employing the classical viscosity coefficient (Model 1) and the other with a viscosity coefficient increased by a factor of ten (Model 2), as motivated by observations. Additionally, we compare these two models across various conditions to assess the influence of other factors on wave behaviors. These factors include thermal conduction, nonlinearity, and transverse structuring of temperature and density within the coronal loop. Through a quantitative comparison of their oscillation $Q$-factors across various scenarios as well as a comparison of wave behaviors across the loop between our simulations and SDO/AIA observations, we have obtained the following results:

1. In Model 1, a flow pulse initiated at the footpoint of a hot and dense loop to simulate the flare heating effect gives rise to a reflected propagating slow wave characterized by a relatively weaker damping rate ($Q\approx2$). Conversely, in Model 2, the same flow pulse leads to the rapid formation of a standing wave within the loop, exhibiting a robust damping rate ($Q\lesssim 1$) that aligns well with observed values. This outcome reinforces the conclusions drawn from 1D MHD modeling conducted by \citet{wan18}.

2. In the nearly isothermal condition ($\gamma=1.05$) assumed in our simulations, the influence of thermal conduction on wave behavior is negligible. The presence of thermal conduction, by means of conductive cooling, counteracts the impact of viscous heating, which would otherwise lead to a gradual rise in the background temperature of the loop if thermal conduction were disabled. The effects of thermal conduction on wave damping in velocity and density perturbations are negligible, although it slightly elevates the damping rate in temperature perturbation.

3. The exclusion of the compressive viscosity term in the energy equation has a negligible impact on the wave behavior in Model 1, but it significantly influences the evolution of loop temperature in Model 2 due to the enhanced viscosity-induced strong viscous heating. However, when compared to the potent dissipation by viscous force in the momentum equation, the impact of viscous heating introduced in the energy equation on wave damping remains negligible. This is probably attributed to the assumption of a near-isothermal conduction.

4. In Model 1 with normal transport coefficients, the onset of standing wave formation within the coronal loop experiences a significant delay due to nonlinearity when subjected to a powerful driving pulse. However, in Model 2, the initiation of standing waves remains almost unaffected by the intensity of the driving pulse. This is because nonlinearity is effectively suppressed by an anomalous increase in viscosity.

5. The transverse density structure within the loop has a limited impact on enhancing the wave trapping. This is indicated by the simulation results, where a hot loop without density contrast shows only a slight decrease in wave amplitude and oscillation $Q$-factor compared to the scenario of a hot and dense loop. In contrast, the transverse temperature structure has a significant impact on the wave behavior, and its effects on Model 1 and Model 2 are notably distinct. In Model 1, for the loop characterized by a lower temperature contrast, the influence of coronal leakage on wave damping outweighs viscous damping. On the other hand, elevating the temperature contrast between the loop and its surroundings promotes wave trapping in the structure, potentially mitigating the rapid escalation of viscous damping with temperature. In Model 2, i.e., with 10 times enhanced viscosity, wave leakage is comparatively insignificant in relation to viscous damping, substantiating the dominant role of viscous damping in this scenario and in qualitative agreement with observations.

6. The analysis of wave evolution across the loop apex reveals notable indications for mode coupling (including the associated propagating fast magnetoacoustic waves and loop kink oscillations) and the pronounced wave leakage due to nonlinear effects in Model 1. In contrast, Model 2 exhibits substantially weaker mode coupling and wave leakage effects. An analysis of three longitudinal loop oscillation events observed by SDO/AIA does not provide clear evidence of mode coupling and wave leakage. The observational features that align with the predictions of Model 2 further support the prior finding of a substantial increase in viscosity. However, our simulations suggest that the expected propagating fast-mode waves generated by mode coupling in Model 1 might be too faint to detect with AIA. Further spectroscopic examination will be necessary in the future. 

Inspired by SOHO/SUMER observations, the rapid excitation of slow-mode standing waves in coronal loops has been investigated by \citet{selw07} using a 2D ideal MHD model in an arcade geometry. They highlighted the significant role of curvature effects in quick formation of standing waves. That is attributed to the curved 2D loop geometry, unlike 1D models, facilitating the interaction of the pulses propagating inside and outside the loop. However, their model assumes a uniform initial gas pressure, leading to the establishment of a dense loop in equilibrium that is cooler than its surrounding corona. This discrepancy contradicts observations that demonstrate the generation of slow waves in hot loops heated by flares. 

In this study of 2.5D MHD modeling, we have established a more realistic hot and dense loop that is initially in equilibrium and remains stable in the entire simulations. This model allows us to investigate more accurately how the loop transverse structuring in density and temperature affects the wave leakage. Moreover, 2.5D models, which have significantly lower computational requirements compared to the 3D case, enable long-term simulations for studying the evolution of slow waves in low-$\beta$ coronal loops. This is especially useful for conducting parametric analyses. On the other hand, a significant limitation of 2.5D models, when compared to 3D models, lies in their simplifications of magnetic geometry and structure. These simplifications constrain their ability to be directly compared with real observations.

In our simulations, we have disregarded the impact of radiative losses on wave damping. It is worth mentioning that the effect of radiative losses on damping of slow magnetosonic waves in coronal loops was considered in a previous 3D MHD study and found to be insignificant \citep{prov18}. In the following, we justify our choice by comparing its characteristic timescale with that of thermal conduction. The significance of thermal conduction in wave dissipation can be assessed through a thermal ratio \citep[see]{dem03}. From Equation 16 in \citet{wan21}, we derive
\begin{equation}
 d=\frac{P_0}{\gamma\tau_{\rm cond}}=\frac{\kappa_0(\gamma-1)(\mu m_p)^{1/2}}{4(\gamma k_B)^{3/2}}\left(\frac{T^2}{nL}\right),
 \label{equ:trt}
\end{equation}
where $P_0=2L/C_s$, and $\tau_{\rm cond}$ is the thermal conduction timescale. Here, $\kappa_0=7.8\times10^{-7}$ is the Spitzer thermal conduction coefficient, $T$ and $n$ are the loop temperature and number density, and $\mu=0.6$.  Similarly, we can define a radiation ratio to quantify the influence of radiative loss on wave damping \citep[see][]{dem04a}. From Equation 37 in \citet{wan21}, we have 
\begin{equation}
 r=\frac{P_0}{\tau_{\rm rad}} =\frac{C(\gamma-1)(\mu m_p)^{1/2}}{(\gamma k_B)^{3/2}}\left(\frac{nL}{T^{13/6}}\right),
\label{equ:rrt}
\end{equation}
where $\tau_{\rm rad}$ is the radiation timescale, and $C=1.86\times10^{-18}$ is a coefficient in the radiative loss function $\Lambda(T)=CT^{-2/3}$ for $T\approx2-10$ MK \citep{rosn78}. 

For the loop model with $T=1.5T_0=10.5$ MK, $n=1.5n_0=1.5\times10^9$ cm$^{-3}$, $L=187$ Mm, and $\gamma=1.05$ in this study, we estimate $d$=0.022 and $\tau_{\rm cond}/P_0$=43 using Equation~\ref{equ:trt}. The prediction of $\tau_{\rm cond}\gg P_0$ aligns with our simulation result that thermal conduction dissipation is negligible in a near-isothermal condition. Using Equation~\ref{equ:rrt} with the same physical parameters, we estimate $r=9.2\times10^{-4}$ and $\tau_{\rm rad}/P_0$=1087. Thus we find $\tau_{\rm rad}/\tau_{\rm cond}$=25. This implies that wave dissipation due to radiative loss is much weak compared to thermal conduction in hot flare plasmas.

Recent 3D MHD simulations by \citet{ofm22} suggested that higher transverse temperature structures in coronal loops can lead to wave trapping by forming a leaky waveguide inside the loop. However, their conclusion carries significant uncertainties as it was based on a qualitative comparison with a previous 3D model with different initial setups and physical parameters in \citet{ofm12}. Additionally, they did not address how damping is affected by the competition between wave trapping and viscous dissipation depending on the loop temperature contrast. 

In our study, through quantitative comparisons of three cases with different temperature contrasts ($\chi_T$=1, 1.5, and 2), we demonstrate that increasing the temperature contrast of the loop to the ambient plasma substantially enhances wave trapping. This effect mitigates the impact of viscous damping, which is known to depend on temperature. For the hotter loop with a temperature contrast of around 2, our simulations indicate that the damping rate resulting from wave leakage is of a similar magnitude to that caused by compressive viscosity with the classical coefficient. However, the damping of simulated oscillations is much weaker compared to the observations in \citet{wan15} even though we have taken coronal leakage into account in the 2D model. This suggests that invoking enhanced viscosity, as proposed in the 1D model by \citet{wan18}, is still necessary. The enhanced compressive viscosity not only facilitate the quick formation of standing waves but also suppress the nonlinear effect in impulsively-generated waves with large amplitudes. This suppression occurs by efficiently dissipating high-frequency components, as suggested by the 1D modeling in \citet{wan18,wan19} and the 2D case studied here. Alternatively, it is possible that coronal leakage in the 3D case could be more significant than in the 2D case as demonstrated in \citet{ofm22}.

Furthermore, when we compare the wave behaviors across the loop in our simulations with observations of several slow wave events using SDO/AIA, it becomes evidence that these observations lack clear indications of mode coupling and wave leakage effects. This finding provides additional substantiation for our conclusion regarding the significant increase in compressive viscosity in flare-heated hot plasma, which can effectively dampen nonlinearity. It is worth noting that a 2.5D model, including the chromosphere and considering thermal conduction only, was employed by \citet{fang15}. Their findings indicate no wave leakage across the loop during the impact of slow waves on the footpoints. This outcome contradicts the predictions made by Model 1 in our simulations. 

There could be several reasons contributing to this disparity: 

1. Temperature contrast: In the model of \citet{fang15}, an initial thermal conduction front rapidly heats the loop to a high temperature of around 10 MK, creating a substantial temperature contrast of 4 compared to the background corona at about 2 MK. This results in an effective waveguide that can trap the waves efficiently. In contrast, our approach assumes a hot loop with a maximum temperature contrast of 1.5 and a background temperature of 7 MK, leading to relatively more significant leakage of waves into the corona.

2. Role of thermal conduction: In \citet{fang15}, the simulated waves dissipate rapidly due to strong thermal conduction, resulting in weak nonlinear effects. In our models, we assume a near-isothermal condition in which thermal conduction plays a less significant role in wave dissipation.

3. Observational consistency: The simulation conducted by \citet{fang15} indicates that the excited reflected slow waves decay out within approximately two periods in the synthesized emissions in both the AIA 131 and 94 \AA\ channels. This is due to the rapid cooling of the heated loop, which falls below the temperature-sensitive ranges of the channels. This is inconsistent with actual observations, where the waves were observed transitioning from the propagating phase into the standing phase lasting over 6 cycles in the 94 \AA\ channel, as reported by \citet{kri21}. This suggests that real observations may not exhibit as strong thermal conduction as theory-predicted, implying some degree of thermal conduction suppression. Enhanced viscosity, as indicated by observations \citep{wan15} and Model 2 in our study, may be a significant factor in wave damping and excitation, as well as in mitigating nonlinear effects .        

Distinct from the SUMER-observed hot loop oscillations that are mostly interpreted as standing waves, the AIA-observed longitudinal oscillations, characterized by sloshing motions, have mostly been interpreted as the reflected propagating slow waves. Based on the observed fact that the damping depends on the oscillation amplitude, \citet{nak19} suggested that the expected enhanced dissipation of higher harmonics could be counteracted by nonlinearity, so enabling the persistence of wave pulses with a sustained shape.

A comparison of simulated waves between Cases 1 and 1B provides evidence that appears to corroborate this scenario. Our simulation results indicate that, for the excited waves with small amplitudes in Case 1B, a standing wave forms more quickly compared to Case 1, where the waves have large amplitudes. This result suggests that nonlinear interactions become significant in propagating wave pulses with substantial amplitudes. These interactions initiate a nonlinear cascade \citep[e.g.,][]{afa15, nak17}, which can postpone the conversion of a propagating wave pulse into a standing wave. This delay occurs because energy is consistently redistributed from large scales to smaller scales. This redistribution can lead to the generation of higher harmonics, which take time to dissipate effectively. Consequently, the process of converting a propagating wave into a standing wave may experience slowdown, influenced by the interplay between nonlinear cascade and dissipation effects. 

The favorable conditions for the excitation of either a propagating or standing wave may be linked to the competition between two processes: viscous dissipation and nonlinear cascade. When dissipation prevails over the nonlinear cascade, as observed in Case 2 due to the presence of anomalous compressive viscosity, a standing wave is quickly excited. Conversely, when the nonlinear cascade outpaces dissipation, as seen in Case 1 for waves with large amplitudes under conditions featuring classical transport coefficients, the excitation of a propagating wave occurs, or its transformation into the standing mode is significantly delayed. To address the discrepancies in the interpretations of slow-mode oscillations observed by SUMER and AIA, particularly in identifying the wave modes as either propagating or standing, a comprehensive statistic study of flare-induced slow wave events is necessary. Such an analysis could utilize simultaneous imaging and spectroscopic observations, which could be made possible by future missions like the {\it Multi-Slit Solar Explorer} (MUSE) \citep{dep20} and the {\it EUV High-Throughput Spectroscopic Telescope} (Solar-C/EUVST) \citep{shim20}. Further research, combined with future observational data, will continue to enhance our understanding of the intricate interplay between wave phenomena and nonlinear processes. This will shed light on their implications for coronal heating processes and further advance the field of coronal seismology.

%%%%%%%%%%%%%%%%%%%%%%%%%%%%%%%%%%%%%%%%%%%%%%%%%%%%%%%%%%%%%%%%%%%%%%%%%%%
%% Acknowledgements
%
  \begin{acks}
SDO is a mission of NASA’s Living With A Star program, and AIA is an instrument onboard SDO, developed and constructed by LMSAL. Resources supporting this work were provided by the NASA High-End Computing (HEC) Program through the NASA Advanced Supercomputing (NAS) Division at Ames Research Center.
  \end{acks}

 \begin{authorcontribution}
T.W. constructed the initial setup of the arcade AR models, conducted all simulation work across various cases, analyzed the simulation and observed data, and measured the parameters listed in tables, prepared all figures and tables, and wrote the manuscript. L.O. developed the 3D nonlinear MHD code NLRAT, provided instructions on its usage for modeling, and contributed to the analysis of the results. L.O. and S.J.B. reviewed the manuscript and offered valuable suggestions and comments.
\end{authorcontribution}

\begin{fundinginformation}
The authors acknowledge support by NASA grant 80NSSC22K0755. TW and LO also acknowledge support from NASA grants 80NSSC18K1131 and 80NSSC21K1687, as well as from NASA GSFC through Cooperative Agreement 80NSSC21M0180 to Catholic University of America, Partnership for Heliophysics and Space Environment Research (PHaSER). 
\end{fundinginformation}

 \begin{ethics}
 \begin{conflict}
The authors declare that they have no conflicts of interest. 
 \end{conflict}
 \end{ethics}

%%% %%%%%%%%%%%%%%%%%%%%%%%%%%%%%%%%%%%%%%%%%%%%%%%%%%%%%%%%%%%
%% Bibliography
%
% Using BibTeX
%
\bibliographystyle{spr-mp-sola}
\bibliography{wang_sola}

\end{article} 
\end{document}